%%%%%%%%%%%%%%%%%%%%%%%%%%%%%%%%%%%%%%%%%%%%%%%%%%%%%%%%%%%%%%%%%%%%%
%
%
%                                      Andrea PASQUINUCCI, 1988
%              PANDA.TEX               S.I.S.S.A., Trieste, Italy
%                                      (Revised 1991, Princeton, USA)
%
%--------------------------------------------------------------------
%
%    These are TEX macros. They work with PLAIN TEX (the basis
%    version of TEX). The only problem can be with the double-page
%    format since it depends on the type of software and laserwriter
%    you use to print, so I cannot guarantee that the double-page
%    format will work properly. Double-page MUST be printed in
%    LANDSCAPE orientation. (You shouldn't have troubles with fonts;
%    if you do, please let me know.)
%
%--------------------------------------------------------------------
%
%                     INTERACTIVE SECTION
%
%--------------------------------------------------------------------
%
\def\standardrisposta{s }\def\reducedrisposta{r }
\def\mplarisposta{mpla }\def\zerorisposta{z }
\def\doublerisposta{d }\def\cartarisposta{e }\def\amsrisposta{y }
\newcount\ingrandimento \newcount\sinnota \newcount\dimnota
\newcount\unoduecol \newdimen\collhsize \newdimen\tothsize
\newdimen\fullhsize \newcount\controllorisposta \sinnota=1
\newskip\infralinea  \global\controllorisposta=0
\immediate\write16 { ********  Welcome to PANDA macros (Plain TeX,
AP, 1991) ******** }
%\immediate\write16 { You'll have to answer a few questions in
%lowercase.}
%\message{>  Do you want it in double-page (d), reduced (r)
%or standard format (s) ? }\read-1 to\risposta
%
%\message{>  Do you want it in USA A4 (u) or EUROPEAN A4
%(e) paper size ? }\read-1 to\srisposta
%
%\message{>  Do you have AMSFonts 2.0 (math) fonts (y/n) ? }
%\read-1 to\arisposta
%
%--------------------------------------------------------------------
%
%             END INTERACTIVE SECTION - PAGE FORMATTING
%
%--------------------------------------------------------------------
%       The following parameters define defaults to the interactive
%       session.  At the moment I have set EUROPEAN and MATH FONTS
%
\def\risposta{s } 
\def\srisposta{e } 
\def\arisposta{y }
\ifx\risposta\standardrisposta \ingrandimento=1200
\message {>> This will come out UNREDUCED << }
\dimnota=2 \unoduecol=1 \global\controllorisposta=1 \fi
\ifx\risposta\reducedrisposta \ingrandimento=1095 \dimnota=1
\unoduecol=1  \global\controllorisposta=1
\message {>> This will come out REDUCED << } \fi
\ifx\risposta\doublerisposta \ingrandimento=1000 \dimnota=2
\unoduecol=2

\message {>> You must print this in
LANDSCAPE orientation << } \global\controllorisposta=1 \fi
\ifx\risposta\mplarisposta \ingrandimento=1000 \dimnota=1
\message {>> Mod. Phys. Lett. A format << }
\unoduecol=1 \global\controllorisposta=1 \fi
\ifx\risposta\zerorisposta \ingrandimento=1000 \dimnota=2
\message {>> Zero Magnification format << }
\unoduecol=1 \global\controllorisposta=1 \fi
\ifnum\controllorisposta=0  \ingrandimento=1200
\message {>>> ERROR IN INPUT, I ASSUME STANDARD
UNREDUCED FORMAT <<< }  \dimnota=2 \unoduecol=1 \fi
\magnification=\ingrandimento
%
%--------------------------------------------------------------------
%
%                        PARAMETERS SETTING
%
%  You can modify these parameters at your will (and resposability)
%--------------------------------------------------------------------
%
\newdimen\eucolumnsize \newdimen\eudoublehsize \newdimen\eudoublevsize
\newdimen\uscolumnsize \newdimen\usdoublehsize \newdimen\usdoublevsize
\newdimen\eusinglehsize \newdimen\eusinglevsize \newdimen\ussinglehsize
\newskip\standardbaselineskip \newdimen\ussinglevsize
\newskip\reducedbaselineskip \newskip\doublebaselineskip
\eucolumnsize=12.0truecm    % column h-size for european doublepage
                            % (12.0treucm default)
\eudoublehsize=25.5truecm   % sheet h-size for european duoblepage
                            % (25.5treucm default)
\eudoublevsize=6.7truein    % sheet v-size for european doublepage
                            % (6.5treuin default  or 17truecm?)
\uscolumnsize=4.4truein     % column h-size for american doublepage
                            % (4.4treuin default)
\usdoublehsize=9.4truein    % sheet h-size for american duoblepage
                            % (9.4treuin default)
\usdoublevsize=6.8truein    % sheet v-size for american doublepage
                            % (6.8treuin default)
\eusinglehsize=6.5truein    % sheet h-size for european singlepage
                            % (6.5truein default)
\eusinglevsize=24truecm     % sheet v-size for european singlepage
                            % (24truecm default)
\ussinglehsize=6.5truein    % sheet h-size for american singlepage
                            % (6.5truein default)
\ussinglevsize=8.9truein    % sheet v-size for american singlepage
                            % (8.9truein default)
\standardbaselineskip=16pt plus.2pt  % baselineskip for standard
                                     % format (16pt default)
\reducedbaselineskip=14pt plus.2pt   % baselineskip for reduced
                                     % format (14pt default)
\doublebaselineskip=12pt plus.2pt    % baselineskip for doublepage
                                     % format (12pt default)
%
%  \Portoffset and \Landoffset define the horizontal and vertical
%  offsets respectively for portrait and landscape modes. Example:
%  \def\Portoffset{\voffset=.4truein\hoffset=.125truein}
%
\def\Portoffset{}
\def\Landoffset{\voffset=-.2truein}
\ifx\risposta\mplarisposta \def\Portoffset{\hoffset=1.8truecm} \fi
%
%  \Landspec defines the \special command that sets the printer
%  to landscape mode without need to specify it directly in the
%  TeX to postscript translator (the command is site dependent).
%  Example: \def\Landspec{\special{ps: landscape}}
%
\def\Landspec{}
\tolerance=10000
\parskip=0pt plus2pt  \leftskip=0pt \rightskip=0pt
%
%   Do not modify anything of what follows
%                       (unless you know what you are doing!)
%----------------------------------------------------------------------
%
\ifx\risposta\standardrisposta \infralinea=\standardbaselineskip \fi
\ifx\risposta\reducedrisposta  \infralinea=\reducedbaselineskip \fi
\ifx\risposta\doublerisposta   \infralinea=\doublebaselineskip \fi
\ifx\risposta\mplarisposta     \infralinea=13pt \fi
\ifx\risposta\zerorisposta     \infralinea=12pt plus.2pt\fi
\ifnum\controllorisposta=0    \infralinea=\standardbaselineskip \fi
\ifx\risposta\doublerisposta   \Landoffset \else \Portoffset \fi
\ifx\risposta\doublerisposta \ifx\srisposta\cartarisposta
\tothsize=\eudoublehsize \collhsize=\eucolumnsize
\vsize=\eudoublevsize  \else  \tothsize=\usdoublehsize
\collhsize=\uscolumnsize \vsize=\usdoublevsize \fi \else
\ifx\srisposta\cartarisposta \tothsize=\eusinglehsize
\vsize=\eusinglevsize \else  \tothsize=\ussinglehsize
\vsize=\ussinglevsize \fi \collhsize=4.4truein \fi
\ifx\risposta\mplarisposta \tothsize=5.0truein
\vsize=7.8truein \collhsize=4.4truein \fi
%
%--------------------------------------------------------------------
%
%                            FONTS
%
%--------------------------------------------------------------------
%
\newcount\contaeuler \newcount\contacyrill \newcount\contaams
\font\ninerm=cmr9  \font\eightrm=cmr8  \font\sixrm=cmr6
\font\ninei=cmmi9  \font\eighti=cmmi8  \font\sixi=cmmi6
\font\ninesy=cmsy9  \font\eightsy=cmsy8  \font\sixsy=cmsy6
\font\ninebf=cmbx9  \font\eightbf=cmbx8  \font\sixbf=cmbx6
\font\ninett=cmtt9  \font\eighttt=cmtt8  \font\nineit=cmti9
\font\eightit=cmti8 \font\ninesl=cmsl9  \font\eightsl=cmsl8
\skewchar\ninei='177 \skewchar\eighti='177 \skewchar\sixi='177
\skewchar\ninesy='60 \skewchar\eightsy='60 \skewchar\sixsy='60
\hyphenchar\ninett=-1 \hyphenchar\eighttt=-1 \hyphenchar\tentt=-1
\def\bfmath{\cmmib}                 % math italic bold \bfmath
\font\tencmmib=cmmib10  \newfam\cmmibfam  \skewchar\tencmmib='177
                  % math bold (cal) symbols
\font\tencmbsy=cmbsy10  \newfam\cmbsyfam  \skewchar\tencmbsy='60
\def\scaps{\cmcsc}                 % small caps (uppercase)
\font\tencmcsc=cmcsc10  \newfam\cmcscfam
\ifnum\ingrandimento=1095

\font\capsone=cmcsc10 at 10.95pt \font\capstwo=cmcsc10 at 13.145pt

\else

\font\capsone=cmcsc10 at 12pt \font\capstwo=cmcsc10 at 14.4pt
\fi

\def\ttaarr{\bf}		% chapter titles' font
\def\ppaarr{\sl}		% section titles' font

%
     % inch-high caps (enormous)
%
%   AMS fonts (this works only if you have at least the 2.0
%              version of AMSFonts, otherwise say no)
%
\newfam\eufmfam \newfam\msamfam \newfam\msbmfam \newfam\eufbfam
\def\Loadeulerfonts{\global\contaeuler=1 \ifx\arisposta\amsrisposta
\font\teneufm=eufm10              %  \eufm   Gothic (or Euler)
\font\eighteufm=eufm8 \font\nineeufm=eufm9 \font\sixeufm=eufm6
\font\seveneufm=eufm7  \font\fiveeufm=eufm5
\font\teneufb=eufb10              %  \eufb   Bold Gothic (or Euler)
\font\eighteufb=eufb8 \font\nineeufb=eufb9 \font\sixeufb=eufb6
\font\seveneufb=eufb7  \font\fiveeufb=eufb5
\font\teneurm=eurm10              %  \eurm   Roman Gothic (or Euler)
\font\eighteurm=eurm8 \font\nineeurm=eurm9
\font\teneurb=eurb10              %  \eurb   Roman Bold Gothic
\font\eighteurb=eurb8 \font\nineeurb=eurb9
\font\teneusm=eusm10              %  \eusm   Slanted Capital Gothic
\font\eighteusm=eusm8 \font\nineeusm=eusm9
\font\teneusb=eusb10              %\eusb Slanted Capital Bold Gothic
\font\eighteusb=eusb8 \font\nineeusb=eusb9
\else \def\eufm{\tt} \def\eufb{\tt} \def\eurm{\tt} \def\eurb{\tt}
\def\eusm{\tt} \def\eusb{\tt}    \fi}
\def\loadeuler{\Loadeulerfonts\tenpoint}
\def\loadamsmath{\global\contaams=1 \ifx\arisposta\amsrisposta
\font\tenmsam=msam10 \font\ninemsam=msam9 \font\eightmsam=msam8
\font\sevenmsam=msam7 \font\sixmsam=msam6 \font\fivemsam=msam5
\font\tenmsbm=msbm10 \font\ninemsbm=msbm9 \font\eightmsbm=msbm8
\font\sevenmsbm=msbm7 \font\sixmsbm=msbm6 \font\fivemsbm=msbm5
\else \def\msbm{\bf} \fi \def\Bbb{\msbm} \def\symbl{\msam} \tenpoint}
\def\loadcyrill{\global\contacyrill=1 \ifx\arisposta\amsrisposta
\font\tenwncyr=wncyr10 \font\ninewncyr=wncyr9 \font\eightwncyr=wncyr8
\font\tenwncyb=wncyr10 \font\ninewncyb=wncyr9 \font\eightwncyb=wncyr8
\font\tenwncyi=wncyr10 \font\ninewncyi=wncyr9 \font\eightwncyi=wncyr8
\else \def\cyrill{\sl} \def\cyrilb{\sl} \def\cyrili{\sl} \fi\tenpoint}
\ifx\arisposta\amsrisposta
\font\sevenex=cmex7               %  reduced math symbols
\font\eightex=cmex8  \font\nineex=cmex9
\font\ninecmmib=cmmib9   \font\eightcmmib=cmmib8
\font\sevencmmib=cmmib7 \font\sixcmmib=cmmib6
\font\fivecmmib=cmmib5   \skewchar\ninecmmib='177
\skewchar\eightcmmib='177  \skewchar\sevencmmib='177
\skewchar\sixcmmib='177   \skewchar\fivecmmib='177
\font\ninecmbsy=cmbsy9    \font\eightcmbsy=cmbsy8
\font\sevencmbsy=cmbsy7  \font\sixcmbsy=cmbsy6
\font\fivecmbsy=cmbsy5   \skewchar\ninecmbsy='60
\skewchar\eightcmbsy='60  \skewchar\sevencmbsy='60
\skewchar\sixcmbsy='60    \skewchar\fivecmbsy='60
\font\ninecmcsc=cmcsc9    \font\eightcmcsc=cmcsc8     \else
\def\cmmib{\fam\cmmibfam\tencmmib}\textfont\cmmibfam=\tencmmib
\scriptfont\cmmibfam=\tencmmib \scriptscriptfont\cmmibfam=\tencmmib
\def\cmbsy{\fam\cmbsyfam\tencmbsy} \textfont\cmbsyfam=\tencmbsy
\scriptfont\cmbsyfam=\tencmbsy \scriptscriptfont\cmbsyfam=\tencmbsy
\scriptfont\cmcscfam=\tencmcsc \scriptscriptfont\cmcscfam=\tencmcsc
\def\cmcsc{\fam\cmcscfam\tencmcsc} \textfont\cmcscfam=\tencmcsc \fi
\catcode`@=11
\newskip\ttglue
\gdef\tenpoint{\def\rm{\fam0\tenrm}
  \textfont0=\tenrm \scriptfont0=\sevenrm \scriptscriptfont0=\fiverm
  \textfont1=\teni \scriptfont1=\seveni \scriptscriptfont1=\fivei
  \textfont2=\tensy \scriptfont2=\sevensy \scriptscriptfont2=\fivesy
  \textfont3=\tenex \scriptfont3=\tenex \scriptscriptfont3=\tenex
  \def\mcal{\fam2 \tensy}  \def\mmit{\fam1 \teni}
  \textfont\itfam=\tenit \def\it{\fam\itfam\tenit}
  \textfont\slfam=\tensl \def\sl{\fam\slfam\tensl}
  \textfont\ttfam=\tentt \scriptfont\ttfam=\eighttt
  \scriptscriptfont\ttfam=\eighttt  \def\tt{\fam\ttfam\tentt}
  \textfont\bffam=\tenbf \scriptfont\bffam=\sevenbf
  \scriptscriptfont\bffam=\fivebf \def\bf{\fam\bffam\tenbf}
     \ifx\arisposta\amsrisposta    \ifnum\contaeuler=1
  \textfont\eufmfam=\teneufm \scriptfont\eufmfam=\seveneufm
  \scriptscriptfont\eufmfam=\fiveeufm \def\eufm{\fam\eufmfam\teneufm}
  \textfont\eufbfam=\teneufb \scriptfont\eufbfam=\seveneufb
  \scriptscriptfont\eufbfam=\fiveeufb \def\eufb{\fam\eufbfam\teneufb}
  \def\eurm{\teneurm} \def\eurb{\teneurb} \def\eusm{\teneusm}
  \def\eusb{\teneusb}    \fi    \ifnum\contaams=1
  \textfont\msamfam=\tenmsam \scriptfont\msamfam=\sevenmsam
  \scriptscriptfont\msamfam=\fivemsam \def\msam{\fam\msamfam\tenmsam}
  \textfont\msbmfam=\tenmsbm \scriptfont\msbmfam=\sevenmsbm
  \scriptscriptfont\msbmfam=\fivemsbm \def\msbm{\fam\msbmfam\tenmsbm}
     \fi      \ifnum\contacyrill=1     \def\cyrill{\tenwncyr}
  \def\cyrilb{\tenwncyb}  \def\cyrili{\tenwncyi}         \fi
  \textfont3=\tenex \scriptfont3=\sevenex \scriptscriptfont3=\sevenex
  \def\cmmib{\fam\cmmibfam\tencmmib} \scriptfont\cmmibfam=\sevencmmib
  \textfont\cmmibfam=\tencmmib  \scriptscriptfont\cmmibfam=\fivecmmib
  \def\cmbsy{\fam\cmbsyfam\tencmbsy} \scriptfont\cmbsyfam=\sevencmbsy
  \textfont\cmbsyfam=\tencmbsy  \scriptscriptfont\cmbsyfam=\fivecmbsy
  \def\cmcsc{\fam\cmcscfam\tencmcsc} \scriptfont\cmcscfam=\eightcmcsc
  \textfont\cmcscfam=\tencmcsc \scriptscriptfont\cmcscfam=\eightcmcsc
     \fi            \tt \ttglue=.5em plus.25em minus.15em
  \normalbaselineskip=12pt
  \setbox\strutbox=\hbox{\vrule height8.5pt depth3.5pt width0pt}
  \let\sc=\eightrm \let\big=\tenbig   \normalbaselines
  \baselineskip=\infralinea  \rm}
\gdef\ninepoint{\def\rm{\fam0\ninerm}
  \textfont0=\ninerm \scriptfont0=\sixrm \scriptscriptfont0=\fiverm
  \textfont1=\ninei \scriptfont1=\sixi \scriptscriptfont1=\fivei
  \textfont2=\ninesy \scriptfont2=\sixsy \scriptscriptfont2=\fivesy
  \textfont3=\tenex \scriptfont3=\tenex \scriptscriptfont3=\tenex
  \def\mcal{\fam2 \ninesy}  \def\mmit{\fam1 \ninei}
  \textfont\itfam=\nineit \def\it{\fam\itfam\nineit}
  \textfont\slfam=\ninesl \def\sl{\fam\slfam\ninesl}
  \textfont\ttfam=\ninett \scriptfont\ttfam=\eighttt
  \scriptscriptfont\ttfam=\eighttt \def\tt{\fam\ttfam\ninett}
  \textfont\bffam=\ninebf \scriptfont\bffam=\sixbf
  \scriptscriptfont\bffam=\fivebf \def\bf{\fam\bffam\ninebf}
     \ifx\arisposta\amsrisposta  \ifnum\contaeuler=1
  \textfont\eufmfam=\nineeufm \scriptfont\eufmfam=\sixeufm
  \scriptscriptfont\eufmfam=\fiveeufm \def\eufm{\fam\eufmfam\nineeufm}
  \textfont\eufbfam=\nineeufb \scriptfont\eufbfam=\sixeufb
  \scriptscriptfont\eufbfam=\fiveeufb \def\eufb{\fam\eufbfam\nineeufb}
  \def\eurm{\nineeurm} \def\eurb{\nineeurb} \def\eusm{\nineeusm}
  \def\eusb{\nineeusb}     \fi   \ifnum\contaams=1
  \textfont\msamfam=\ninemsam \scriptfont\msamfam=\sixmsam
  \scriptscriptfont\msamfam=\fivemsam \def\msam{\fam\msamfam\ninemsam}
  \textfont\msbmfam=\ninemsbm \scriptfont\msbmfam=\sixmsbm
  \scriptscriptfont\msbmfam=\fivemsbm \def\msbm{\fam\msbmfam\ninemsbm}
     \fi       \ifnum\contacyrill=1     \def\cyrill{\ninewncyr}
  \def\cyrilb{\ninewncyb}  \def\cyrili{\ninewncyi}         \fi
  \textfont3=\nineex \scriptfont3=\sevenex \scriptscriptfont3=\sevenex
  \def\cmmib{\fam\cmmibfam\ninecmmib}  \textfont\cmmibfam=\ninecmmib
  \scriptfont\cmmibfam=\sixcmmib \scriptscriptfont\cmmibfam=\fivecmmib
  \def\cmbsy{\fam\cmbsyfam\ninecmbsy}  \textfont\cmbsyfam=\ninecmbsy
  \scriptfont\cmbsyfam=\sixcmbsy \scriptscriptfont\cmbsyfam=\fivecmbsy
  \def\cmcsc{\fam\cmcscfam\ninecmcsc} \scriptfont\cmcscfam=\eightcmcsc
  \textfont\cmcscfam=\ninecmcsc \scriptscriptfont\cmcscfam=\eightcmcsc
     \fi            \tt \ttglue=.5em plus.25em minus.15em
  \normalbaselineskip=11pt
  \setbox\strutbox=\hbox{\vrule height8pt depth3pt width0pt}
  \let\sc=\sevenrm \let\big=\ninebig \normalbaselines\rm}
\gdef\eightpoint{\def\rm{\fam0\eightrm}
  \textfont0=\eightrm \scriptfont0=\sixrm \scriptscriptfont0=\fiverm
  \textfont1=\eighti \scriptfont1=\sixi \scriptscriptfont1=\fivei
  \textfont2=\eightsy \scriptfont2=\sixsy \scriptscriptfont2=\fivesy
  \textfont3=\tenex \scriptfont3=\tenex \scriptscriptfont3=\tenex
  \def\mcal{\fam2 \eightsy}  \def\mmit{\fam1 \eighti}
  \textfont\itfam=\eightit \def\it{\fam\itfam\eightit}
  \textfont\slfam=\eightsl \def\sl{\fam\slfam\eightsl}
  \textfont\ttfam=\eighttt \scriptfont\ttfam=\eighttt
  \scriptscriptfont\ttfam=\eighttt \def\tt{\fam\ttfam\eighttt}
  \textfont\bffam=\eightbf \scriptfont\bffam=\sixbf
  \scriptscriptfont\bffam=\fivebf \def\bf{\fam\bffam\eightbf}
     \ifx\arisposta\amsrisposta   \ifnum\contaeuler=1
  \textfont\eufmfam=\eighteufm \scriptfont\eufmfam=\sixeufm
  \scriptscriptfont\eufmfam=\fiveeufm \def\eufm{\fam\eufmfam\eighteufm}
  \textfont\eufbfam=\eighteufb \scriptfont\eufbfam=\sixeufb
  \scriptscriptfont\eufbfam=\fiveeufb \def\eufb{\fam\eufbfam\eighteufb}
  \def\eurm{\eighteurm} \def\eurb{\eighteurb} \def\eusm{\eighteusm}
  \def\eusb{\eighteusb}       \fi    \ifnum\contaams=1
  \textfont\msamfam=\eightmsam \scriptfont\msamfam=\sixmsam
  \scriptscriptfont\msamfam=\fivemsam \def\msam{\fam\msamfam\eightmsam}
  \textfont\msbmfam=\eightmsbm \scriptfont\msbmfam=\sixmsbm
  \scriptscriptfont\msbmfam=\fivemsbm \def\msbm{\fam\msbmfam\eightmsbm}
     \fi       \ifnum\contacyrill=1     \def\cyrill{\eightwncyr}
  \def\cyrilb{\eightwncyb}  \def\cyrili{\eightwncyi}         \fi
  \textfont3=\eightex \scriptfont3=\sevenex \scriptscriptfont3=\sevenex
  \def\cmmib{\fam\cmmibfam\eightcmmib}  \textfont\cmmibfam=\eightcmmib
  \scriptfont\cmmibfam=\sixcmmib \scriptscriptfont\cmmibfam=\fivecmmib
  \def\cmbsy{\fam\cmbsyfam\eightcmbsy}  \textfont\cmbsyfam=\eightcmbsy
  \scriptfont\cmbsyfam=\sixcmbsy \scriptscriptfont\cmbsyfam=\fivecmbsy
  \def\cmcsc{\fam\cmcscfam\eightcmcsc} \scriptfont\cmcscfam=\eightcmcsc
  \textfont\cmcscfam=\eightcmcsc \scriptscriptfont\cmcscfam=\eightcmcsc
     \fi             \tt \ttglue=.5em plus.25em minus.15em
  \normalbaselineskip=9pt
  \setbox\strutbox=\hbox{\vrule height7pt depth2pt width0pt}
  \let\sc=\sixrm \let\big=\eightbig \normalbaselines\rm }
\gdef\tenbig#1{{\hbox{$\left#1\vbox to8.5pt{}\right.\n@space$}}}
\gdef\ninebig#1{{\hbox{$\textfont0=\tenrm\textfont2=\tensy
   \left#1\vbox to7.25pt{}\right.\n@space$}}}
\gdef\eightbig#1{{\hbox{$\textfont0=\ninerm\textfont2=\ninesy
   \left#1\vbox to6.5pt{}\right.\n@space$}}}
 %for 10-pt math in 9-pt territory
\def\alternativefont#1#2{\ifx\arisposta\amsrisposta \relax \else
\xdef#1{#2} \fi}
\global\contaeuler=0 \global\contacyrill=0 \global\contaams=0
%
%--------------------------------------------------------------------
%
%                            MACROS
%
%--------------------------------------------------------------------
%
\newbox\fotlinebb \newbox\hedlinebb \newbox\leftcolumn
\gdef\makeheadline{\vbox to 0pt{\vskip-22.5pt
     \fullline{\vbox to8.5pt{}\the\headline}\vss}\nointerlineskip}
\gdef\makehedlinebb{\vbox to 0pt{\vskip-22.5pt
     \fullline{\vbox to8.5pt{}\copy\hedlinebb\hfil
     \line{\hfill\the\headline\hfill}}\vss} \nointerlineskip}
\gdef\makefootline{\baselineskip=24pt \fullline{\the\footline}}
\gdef\makefotlinebb{\baselineskip=24pt
    \fullline{\copy\fotlinebb\hfil\line{\hfill\the\footline\hfill}}}
\gdef\doubleformat{\shipout\vbox{\Landspec\makehedlinebb
     \fullline{\box\leftcolumn\hfil\columnbox}\makefotlinebb}
     \advancepageno}
\gdef\columnbox{\leftline{\pagebody}}
\gdef\line#1{\hbox to\hsize{\hskip\leftskip#1\hskip\rightskip}}
\gdef\fullline#1{\hbox to\fullhsize{\hskip\leftskip{#1}%
\hskip\rightskip}}
\gdef\footnote#1{\let\@sf=\empty
         \ifhmode\edef\#sf{\spacefactor=\the\spacefactor}\/\fi
         #1\@sf\vfootnote{#1}}
\gdef\vfootnote#1{\insert\footins\bgroup
         \ifnum\dimnota=1  \eightpoint\fi
         \ifnum\dimnota=2  \ninepoint\fi
         \ifnum\dimnota=0  \tenpoint\fi
         \interlinepenalty=\interfootnotelinepenalty
         \splittopskip=\ht\strutbox
         \splitmaxdepth=\dp\strutbox \floatingpenalty=20000
         \leftskip=\oldssposta \rightskip=\olddsposta
         \spaceskip=0pt \xspaceskip=0pt
         \ifnum\sinnota=0   \textindent{#1}\fi
         \ifnum\sinnota=1   \item{#1}\fi
         \footstrut\futurelet\next\fo@t}
\gdef\fo@t{\ifcat\bgroup\noexpand\next \let\next\f@@t
             \else\let\next\f@t\fi \next}
\gdef\f@@t{\bgroup\aftergroup\@foot\let\next}
\gdef\f@t#1{#1\@foot} \gdef\@foot{\strut\egroup}
\gdef\footstrut{\vbox to\splittopskip{}}
\skip\footins=\bigskipamount
\count\footins=1000  \dimen\footins=8in
\catcode`@=12
\tenpoint
\ifnum\unoduecol=1 \hsize=\tothsize   \fullhsize=\tothsize \fi
\ifnum\unoduecol=2 \hsize=\collhsize  \fullhsize=\tothsize \fi
\global\let\lrcol=L      \ifnum\unoduecol=1
\output{\plainoutput{\ifnum\tipbnota=2 \clearnmbnota\fi}} \fi
\ifnum\unoduecol=2 \output{\if L\lrcol
     \global\setbox\leftcolumn=\columnbox
     \global\setbox\fotlinebb=\line{\hfill\the\footline\hfill}
     \global\setbox\hedlinebb=\line{\hfill\the\headline\hfill}
     \advancepageno  \global\let\lrcol=R
     \else  \doubleformat \global\let\lrcol=L \fi
     \ifnum\outputpenalty>-20000 \else\dosupereject\fi
     \ifnum\tipbnota=2\clearnmbnota\fi }\fi
\def\ifdoublepage{\ifnum\unoduecol=2 }
\gdef\yespagenumbers{\footline={\hss\tenrm\folio\hss}}
\gdef\ciao{ \ifnum\fdefcontre=1 \endfdef\fi
     \par\vfill\supereject \ifnum\unoduecol=2
     \if R\lrcol  \headline={}\nopagenumbers\null\vfill\eject
     \fi\fi \end}

\newskip\olddsposta \newskip\oldssposta
\global\oldssposta=\leftskip \global\olddsposta=\rightskip

\def\filldots{\leaders\hbox to 1em{\hss.\hss}\hfill}
\def\inquadrb#1 {\vbox {\hrule  \hbox{\vrule \vbox {\vskip .2cm
    \hbox {\ #1\ } \vskip .2cm } \vrule  }  \hrule} }
 \def\newline{\hfil\break}
\def\jump{\vskip\baselineskip} \newskip\iinnffrr
\def\sjump{\iinnffrr=\baselineskip
          \divide\iinnffrr by 2 \vskip\iinnffrr}
\def\bjump{\vskip\baselineskip \vskip\baselineskip}
\newcount\nmbnota  \def\clearnmbnota{\global\nmbnota=0}
\newcount\tipbnota \def\letterfootnote{\global\tipbnota=1}

\def\note#1{\global\advance\nmbnota by 1 \ifnum\tipbnota=1
    \footnote{$^{\rm\nttlett}$}{#1} \else {\ifnum\tipbnota=2
    \footnote{$^{\nttsymb}$}{#1}
    \else\footnote{$^{\the\nmbnota}$}{#1}\fi}\fi}
\def\nttlett{\ifcase\nmbnota \or a\or b\or c\or d\or e\or f\or
g\or h\or i\or j\or k\or l\or m\or n\or o\or p\or q\or r\or
s\or t\or u\or v\or w\or y\or x\or z\fi}
\def\nttsymb{\ifcase\nmbnota \or\dag\or\sharp\or\ddag\or\star\or
\natural\or\flat\or\clubsuit\or\diamondsuit\or\heartsuit
\or\spadesuit\fi}   \clearnmbnota
\def\numberfootnote{\global\tipbnota=0} \numberfootnote
\def\setnote#1{\expandafter\xdef\csname#1\endcsname{
\ifnum\tipbnota=1 {\rm\nttlett} \else {\ifnum\tipbnota=2
{\nttsymb} \else \the\nmbnota\fi}\fi} }
\newcount\nbmfig  \def\clearnbmfig{\global\nbmfig=0}
\gdef\figure{\global\advance\nbmfig by 1
      {\rm fig. \the\nbmfig}}   \clearnbmfig
\def\setfig#1{\expandafter\xdef\csname#1\endcsname{fig. \the\nbmfig}}
 \def\endformula{\eqno\numero $$}
 \def\efr{\endformula}
\newcount\frmcount \def\clearfrmcount{\global\frmcount=0}
\def\numero{\global\advance\frmcount by 1   \ifnum\indappcount=0
  {\ifnum\cpcount <1 {\hbox{\rm (\the\frmcount )}}  \else
  {\hbox{\rm (\the\cpcount .\the\frmcount )}} \fi}  \else
  {\hbox{\rm (\applett .\the\frmcount )}} \fi}
\def\nameformula#1{\global\advance\frmcount by 1%
\ifnum\draftnum=0  {\ifnum\indappcount=0%
{\ifnum\cpcount<1\xdef\spzzttrra{(\the\frmcount )}%
\else\xdef\spzzttrra{(\the\cpcount .\the\frmcount )}\fi}%
\else\xdef\spzzttrra{(\applett .\the\frmcount )}\fi}%
\else\xdef\spzzttrra{(#1)}\fi%
\expandafter\xdef\csname#1\endcsname{\spzzttrra}
\eqno \hbox{\rm\spzzttrra} $$}
\def\nfr{\nameformula}    
\def\nameali#1{\global\advance\frmcount by 1%
\ifnum\draftnum=0  {\ifnum\indappcount=0%
{\ifnum\cpcount<1\xdef\spzzttrra{(\the\frmcount )}%
\else\xdef\spzzttrra{(\the\cpcount .\the\frmcount )}\fi}%
\else\xdef\spzzttrra{(\applett .\the\frmcount )}\fi}%
\else\xdef\spzzttrra{(#1)}\fi%
\expandafter\xdef\csname#1\endcsname{\spzzttrra}
  \hbox{\rm\spzzttrra} }      \clearfrmcount
\newcount\cpcount \def\clearcpcount{\global\cpcount=0}
\newcount\subcpcount \def\clearsubcpcount{\global\subcpcount=0}
\newcount\appcount \def\clearappcount{\global\appcount=0}
\newcount\indappcount \def\clearindappcount{\indappcount=0}
\newcount\sottoparcount 

\def\applett{\ifcase\appcount  \or {A}\or {B}\or {C}\or
{D}\or {E}\or {F}\or {G}\or {H}\or {I}\or {J}\or {K}\or {L}\or
{M}\or {N}\or {O}\or {P}\or {Q}\or {R}\or {S}\or {T}\or {U}\or
{V}\or {W}\or {X}\or {Y}\or {Z}\fi    \ifnum\appcount<0
\immediate\write16 {Panda ERROR - Appendix: counter "appcount"
out of range}\fi  \ifnum\appcount>26  \immediate\write16 {Panda
ERROR - Appendix: counter "appcount" out of range}\fi}
\clearappcount  \clearindappcount \newcount\connttrre
\def\clearconnttrre{\global\connttrre=0} \newcount\countref
\def\clearcountref{\global\countref=0} \clearcountref
\def\chapter#1{\global\advance\cpcount by 1 \clearfrmcount
                 \goodbreak\null\vbox{\jump\nobreak
                 \clearsubcpcount\clearindappcount
                 \itemitem{\ttaarr\the\cpcount .\qquad}{\ttaarr #1}
                 \par\nobreak\jump\sjump}\nobreak}
\def\section#1{\global\advance\subcpcount by 1 \goodbreak\null
               \vbox{\sjump\nobreak\ifnum\indappcount=0
                 {\ifnum\cpcount=0 {\itemitem{\ppaarr
               .\the\subcpcount\quad\enskip\ }{\ppaarr #1}\par} \else
                 {\itemitem{\ppaarr\the\cpcount .\the\subcpcount\quad
                  \enskip\ }{\ppaarr #1} \par}  \fi}
                \else{\itemitem{\ppaarr\applett .\the\subcpcount\quad
                 \enskip\ }{\ppaarr #1}\par}\fi\nobreak\jump}\nobreak}
\clearsubcpcount
\def\appendix#1{\global\advance\appcount by 1 \clearfrmcount
                  \goodbreak\null\vbox{\jump\nobreak
                  \global\advance\indappcount by 1 \clearsubcpcount
          \itemitem{ }{\hskip-40pt\ttaarr #1}
%                  \itemitem{\ttaarr App.\applett\ }{\ttaarr #1}
             \nobreak\jump\sjump}\nobreak}
\clearappcount \clearindappcount
\def\references{\goodbreak\null\vbox{\jump\nobreak
   \noindent{\ttaarr References} \nobreak\jump\sjump}\nobreak}
%   \itemitem{}{\ttaarr References} \nobreak\jump\sjump}\nobreak}

\clearcpcount\clearcountref

\def\setchap#1{\ifnum\indappcount=0{\ifnum\subcpcount=0%
\xdef\spzzttrra{\the\cpcount}%
\else\xdef\spzzttrra{\the\cpcount .\the\subcpcount}\fi}
\else{\ifnum\subcpcount=0 \xdef\spzzttrra{\applett}%
\else\xdef\spzzttrra{\applett .\the\subcpcount}\fi}\fi
\expandafter\xdef\csname#1\endcsname{\spzzttrra}}
\newcount\draftnum \newcount\ppora   \newcount\ppminuti
\global\ppora=\time   \global\ppminuti=\time
\global\divide\ppora by 60  \draftnum=\ppora
\multiply\draftnum by 60    \global\advance\ppminuti by -\draftnum
\def\droggi{\number\day /\number\month /\number\year\ \the\ppora
:\the\ppminuti}     \global\draftnum=0
\def\draftcomment#1{\ifnum\draftnum=0 \relax \else
{\ {\bf ***}\ #1\ {\bf ***}\ }\fi} 
%
%     Maximum number of references = 200
%     boxes 50 -> 250 reserved for references
%
\catcode`@=11
\gdef\Ref#1{\expandafter\ifx\csname @rrxx@#1\endcsname\relax%
{\global\advance\countref by 1    \ifnum\countref>200
\immediate\write16 {Panda ERROR - Ref: maximum number of references
exceeded}  \expandafter\xdef\csname @rrxx@#1\endcsname{0}\else
\expandafter\xdef\csname @rrxx@#1\endcsname{\the\countref}\fi}\fi
\ifnum\draftnum=0 \csname @rrxx@#1\endcsname \else#1\fi}
\gdef\beginref{\ifnum\draftnum=0  \gdef\Rref{\fairef}
\gdef\endref{\scriviref} \else\relax\fi
\ifx\risposta\mplarisposta \ninepoint \fi
\parskip 2pt plus.2pt \baselineskip=12pt}
\def\Reflab#1{[#1]} \gdef\Rref#1#2{\item{\Reflab{#1}}{#2}}
\gdef\endref{\relax}  \newcount\conttemp
\gdef\fairef#1#2{\expandafter\ifx\csname @rrxx@#1\endcsname\relax
{\global\conttemp=0 \immediate\write16 {Panda ERROR - Ref: reference
[#1] undefined}} \else
{\global\conttemp=\csname @rrxx@#1\endcsname } \fi
\global\advance\conttemp by 50  \global\setbox\conttemp=\hbox{#2} }
\gdef\scriviref{\clearconnttrre\conttemp=50
\loop\ifnum\connttrre<\countref \advance\conttemp by 1
\advance\connttrre by 1
\item{\Reflab{\the\connttrre}}{\unhcopy\conttemp} \repeat}
\clearcountref \clearconnttrre
\catcode`@=12
\ifx\risposta\mplarisposta \def\Reflab#1{#1.} \letterfootnote \fi

\def\slashchar#1{\setbox0=\hbox{$#1$} \dimen0=\wd0
     \setbox1=\hbox{/} \dimen1=\wd1 \ifdim\dimen0>\dimen1
      \rlap{\hbox to \dimen0{\hfil/\hfil}} #1 \else
      \rlap{\hbox to \dimen1{\hfil$#1$\hfil}} / \fi}
\ifx\oldchi\undefined \let\oldchi=\chi
  \def\cchi{{\raise 1pt\hbox{$\oldchi$}}} \let\chi=\cchi \fi

\def\frac#1#2{{\textstyle{#1 \over #2}}}

\def\half{\ifinner {\scriptstyle {1 \over 2}}\else {1 \over 2} \fi}

\def\simge{\rlap{\raise 2pt \hbox{$>$}}{\lower 2pt \hbox{$\sim$}}}
\def\simle{\rlap{\raise 2pt \hbox{$<$}}{\lower 2pt \hbox{$\sim$}}}

\def\vbig#1#2{{\vbigd@men=#2\divide\vbigd@men by 2%
\hbox{$\left#1\vbox to \vbigd@men{}\right.\n@space$}}}

%
%--------------------------------------------------------------------
%
\newcount\fdefcontre \newcount\fdefcount \newcount\indcount
\newread\filefdef  \newread\fileftmp  \newwrite\filefdef
\newwrite\fileftmp     \def\strip#1*.A {#1}
\def\futuredef#1{\beginfdef
\expandafter\ifx\csname#1\endcsname\relax%
{\immediate\write\fileftmp {#1*.A}
\immediate\write16 {Panda Warning - fdef: macro "#1" on page
\the\pageno \space undefined}
\ifnum\draftnum=0 \expandafter\xdef\csname#1\endcsname{(?)}
\else \expandafter\xdef\csname#1\endcsname{(#1)} \fi
\global\advance\fdefcount by 1}\fi   \csname#1\endcsname}

\def\beginfdef{\ifnum\fdefcontre=0
\immediate\openin\filefdef \jobname.fdef
\immediate\openout\fileftmp \jobname.ftmp
\global\fdefcontre=1  \ifeof\filefdef \immediate\write16 {Panda
WARNING - fdef: file \jobname.fdef not found, run TeX again}
\else \immediate\read\filefdef to\spzzttrra
\global\advance\fdefcount by \spzzttrra
\indcount=0      \loop\ifnum\indcount<\fdefcount
\advance\indcount by 1   \immediate\read\filefdef to\spezttrra
\immediate\read\filefdef to\sppzttrra
\edef\spzzttrra{\expandafter\strip\spezttrra}
\immediate\write\fileftmp {\spzzttrra *.A}
\expandafter\xdef\csname\spzzttrra\endcsname{\sppzttrra}
\repeat \fi \immediate\closein\filefdef \fi}
\def\endfdef{\immediate\closeout\fileftmp   \ifnum\fdefcount>0
\immediate\openin\fileftmp \jobname.ftmp
\immediate\openout\filefdef \jobname.fdef
\immediate\write\filefdef {\the\fdefcount}   \indcount=0
\loop\ifnum\indcount<\fdefcount    \advance\indcount by 1
\immediate\read\fileftmp to\spezttrra
\edef\spzzttrra{\expandafter\strip\spezttrra}
\immediate\write\filefdef{\spzzttrra *.A}
\edef\spezttrra{\string{\csname\spzzttrra\endcsname\string}}
\iwritel\filefdef{\spezttrra}
\repeat  \immediate\closein\fileftmp \immediate\closeout\filefdef
\immediate\write16 {Panda Warning - fdef: Label(s) may have changed,
re-run TeX to get them right}\fi}
\def\iwritel#1#2{\newlinechar=-1
{\newlinechar=`\ \immediate\write#1{#2}}\newlinechar=-1}
\global\fdefcontre=0 \global\fdefcount=0 \global\indcount=0
%
%--------------------------------------------------------------------
%
\null
%
%--------------------------------------------------------------------
%
%                             THE    END
%
%--------------------------------------------------------------------
%
%\input panda
%\draftmode{Massive}
\loadamsmath
\loadeuler
\mathchardef\bphi="731E
\mathchardef\balpha="710B
\mathchardef\bbeta="710C
\mathchardef\bomega="7121
\mathchardef\btheta="7112
\def\thb{{\bfmath\btheta}}
\def\alb{{\bfmath\balpha}}
\def\beb{{\bfmath\bbeta}}

\def\ss{{\eufm s}}
\def\hh{{\eufm h}}
\pageno=0
\nopagenumbers{\baselineskip=12pt
\line{\hfill US-FT/24-96}
\line{\hfill SWAT/117} 
\line{\hfill\tt hep-th/9606032}
\line{\hfill May 1996}\bjump
\ifdoublepage \bjump\bjump\bjump\else\jump\vfill\fi
\centerline{\capstwo The Symmetric Space and}
\sjump
\centerline{\capstwo Homogeneous sine-Gordon theories}
\bjump\jump
\centerline{{\scaps Carlos R.~Fern\'andez-Pousa$^{\>\dagger}$}, 
{\scaps Manuel~V. Gallas$^{\>\dagger}$},} \sjump
\centerline{{\scaps Timothy J.~Hollowood$^{\>\ddagger}$} and
{\scaps J. Luis Miramontes$^{\>\dagger}$}}
\jump\jump
\centerline{\sl $^{\dagger\>}$Departamento de F\'\i sica de Part\'\i
culas,}
\centerline{\sl Facultad de F\'\i sica,}
\centerline{\sl Universidad de Santiago,}
\centerline{\sl E-15706 Santiago de Compostela, Spain}
\jump
\centerline{\sl $^{\ddagger\>}$Department of Physics,}
\centerline{\sl University of Wales Swansea,}
\centerline{\sl Singleton Park, Swansea SA2 8PP, U.K.}
\jump
\centerline{\tt pousa@gaes.usc.es,~ gallas@fpaxp1.usc.es} \sjump  
\centerline{{\tt t.hollowood@swansea.ac.uk},~ and~ 
{\tt miramont@fpaxp1.usc.es}} 
\bjump\bjump
\ifdoublepage
\vfill {\noindent
\line{May 1996\hfill}}
\eject\null\vfill\fi
\centerline{\capsone ABSTRACT}\jump

\noindent
Two series of integrable theories are constructed 
which have soliton solutions and can be thought of as generalizations 
of the sine-Gordon theory. They exhibit internal symmetries and can be
described as gauged WZW theories with a potential term. The spectrum
of massive states is determined. 
\vfill
\ifdoublepage \else
\noindent
\line{May 1996\hfill}\fi
\eject}
\yespagenumbers\pageno=1
\footline={\hss\tenrm-- \folio\ --\hss}

\chapter{Introduction}

Integrability has proved to be a powerful tool in elucidating
non-perturbative properties of field theory.
For instance, in the sine-Gordon theory an exact quantum
description of the solitons can be deduced [\Ref{ZZ},\Ref{SG}], which can be
shown to match precisely with semi-classical approaches in the
appropriate limit. This work gives us confidence that similar
semi-classical approaches are equally valid in higher dimensions, 
for instance for monopoles in four dimensional
gauge theories. In that
context, though, it is clear that additional complications can arise
that are not captured by the sine-Gordon theory; namely the solitons,
in this case t' Hooft-Polyakov monopoles,
can carry internal quantum numbers and have a complicated spectrum
[\Ref{GENM}]. Therefore, 
it would be useful to have two-dimensional soliton theories
which also have internal symmetries and are open to the same level of
analysis as the sine-Gordon theory. 

Many integrable generalizations of the sine-Gordon
equations-of-motion have already been written down and are known as the
non-abelian affine Toda equations [\Ref{LS},\Ref{NAT},\Ref{LUIZ}]. However, not
all these equations can be derived by extremizing an action with sensible
properties like a positive-definite kinetic term 
and a real potential. In this paper, following on from
[\Ref{NOS}], we show that non-abelian affine Toda equations give rise to two
series of models, referred to as the Symmetric Space Sine-Gordon (SSSG)
theories and the Homogeneous Sine-Gordon (HSG) theories, both of which are
integrable, admit soliton solutions, and have a real positive-definite
action. 

One of the subtleties of these theories is that in general
their potentials have flat directions [\Ref{NOS}]. In order to
construct theories which will admit an S-matrix description, like the
sine-Gordon theory, we will show how the
flat directions in the potential can be removed by
a gauging procedure. The resulting theories are of the form of a 
gauged Wess-Zumino-Witten (WZW) action plus a certain potential which
deforms the theory away from the critical point along an integrable direction.

More specifically, the SSSG theories are related to a compact
symmetric space $G/G_0$,
with a $G_0$-valued field. As a particular example, the ordinary  sine-Gordon
theory corresponds to $G/G_0 = {\rm SU}(2)/{\rm U}(1)$. This class of
theories was first considered in [\Ref{SYMSP}], and their Lagrangian
formulation has been recently worked out in [\Ref{PBAK}]. Nevertheless,
neither the symmetries nor the mass spectrum of these theories have been
investigated before. In fact the theories in~[\Ref{SYMSP},\Ref{PBAK}]
have flat directions which, as we remarked above, can be removed by gauging.
All the models constructed in [\Ref{NOS}] are of this type, with
$G/G_0$ a symmetric space of type~I.

The HSG theories are particular examples of the deformed coset models
constructed in~[\Ref{PARK1}]. The fields of these theory take values in   
some compact semi-simple group $G_0$. It turns out that the
equations-of-motion of the SSSG theories can be obtained by reduction
from an appropriate HSG model.
The simplest representative of this class is the complex sine-Gordon
theory~[\Ref{PARK1},\Ref{BAKAS},\Ref{CSG},\Ref{PARK2}] for which 
$G_0= {\rm SU}(2)$. In fact, all the HSG theories can be understood as
interacting sets of
complex sine-Gordon fields.\note{A complete discussion of the HSG
theories will appear in [\Ref{COMING}].} 

The overall aim of
the analysis is to completely determine the quantum properties of these
theories by constructing a factorizable S-matrix which describes the
solitons and particle states of the theory, in the same way that the
sine-Gordon theory has been described [\Ref{ZZ},\Ref{SG}].
In the present paper, our
aims will be more modest, and we shall restrict ourselves to
an explanation of how these theories are formulated
at the Lagrangian level. The most important issue that we tackle, is
to show how the flat directions of the
potential can be removed by a gauging procedure. Actually, the condition that
this procedure succeeds in removing all the flat directions requires
that the theories are related to a coset 
$G_0/U(1)^{\times p}$. In other words, even though the field will be
non-abelian, the gauge and global symmetries of these theories will always be
abelian. In contrast to~[\Ref{PBAK},\Ref{PARK1}], we shall allow for more
general types of gauging over, and above, the vector type. We will then
derive the spectrum of particle states corresponding to small fluctuations
around the vacuum. Finally, we will go on to explain briefly how soliton
solutions can be found by the Leznov-Saveliev procedure and how  soliton
masses and classical scattering time-delays can be found.

\chapter{Massive theories and non-abelian affine Toda equations}

The quantum field theories that we will construct are characterized
by the fact that their equations-of-motion are 
non-abelian affine Toda equations~[\Ref{LS},\Ref{NAT},\Ref{LUIZ}]. 
These equations are usually written down using affine (Kac-Moody)
algebras, but for our purposes, it will be more convenient to describe them
directly by means of finite Lie algebras.

In order to keep the construction as general as possible, let $g$ be a
complex semi-simple finite-dimensional Lie algebra, and 
$\sigma$ a finite order automorphism of $g$.\note{On
$g$, we denote the invariant and non-degenerate Killing form 
by $\langle\;,\; \rangle$, normalized such that long roots have
square length $2$.} If $N$ is the order of $\sigma$, $\sigma^N =I$, it
induces a ${\Bbb Z}/N{\Bbb Z}$ gradation of $g$:
$$
g \> =\> \bigoplus_{j\in{\Bbb Z}} g_{\overline \jmath}\>, \quad
[g_{\overline \jmath}\>,
\>g_{\overline k}]
\> \subset \> g_{\overline{j+k}} \>,
\efr
where $ \sigma(a)\> =\> \exp(2\pi ij/N)a$, for any $a\in g_{\overline
\jmath}$,  and ${\overline \jmath}$
stands for the residue of $j$ mod $N$. 
If $g$ is simple, then let $r=1,2$ or $3$ be the 
least positive integer such that $\sigma^r$ is an inner automorphism
of $g$. In this case, $(g,\sigma)$ provide a realization
of the (twisted) affine Kac-Moody algebra $g^{(r)}$ in terms of the
central extension of the (twisted) loop algebra
$$
{\cal L}(g, \sigma)\> = \> \bigoplus_{j\in{\Bbb Z}} z^j\otimes
g_{\overline \jmath}\>.
\efr

The fields of the theory, $h(x,t)$, 
take values in the group $G_0$ associated to
the Lie algebra of the zero-graded component $g_{\overline 0}$, and
the Toda equation involves the choice of two ad-diagonalizable
(semi-simple) elements $\Lambda_+\in g_{\overline k}$ and $\Lambda_-\in
g_{\overline{ N-k}}$, for some non-negative integer number $k$. 
The Toda equations are
$$
\partial_-\left(h^{-1}\partial_+h\right)\> =\> -\> m^2 \left[\Lambda_+\>
,\> h^{-1} \Lambda_- h\right]
\nfr{NaT}
or equivalently
$$
\partial_+\left(\partial_-h h^{-1}\right)\> =\> -\>m^2 \left[h\>
\Lambda_+ \>h^{-1}\> ,\> \Lambda_- \right],
\nfr{NaTO}
where $x_\pm = t \pm x$ are the light-cone variables and $m$ is a
constant with dimensions of mass. 
The integrability of the Toda equations
is manifested in the equivalent zero-curvature form
$$
\left[\partial_+ \> +\> h^{-1}\partial_+h \> + \>im\Lambda_+ \>,\>
\partial_-\> +\>im h^{-1} \Lambda_- h\right]\> =\>0 \>.
\efr

The equations~\NaT\ follow from extremizing the action
$$
S[h]\>=\> {1 \over \beta^2} \biggl\{ S_{\rm WZW}[h] \>- \> \int
d^2 x \>V(h) \biggl\}\>,
\nfr{Act}
where the kinetic term $S_{\rm WZW}[h]$ is the
Wess-Zumino-Witten action for the group $G_0$, and the potential is  
$$
V(h)\> =\> - \> {m^2 \over 2\pi}\> \langle  \Lambda_+\>,\>  h^{-1}\>
\Lambda_- \> h\rangle\>.
\nfr{Pot}
In~\Act, $\beta$ is a coupling constant that plays no role in the
classical theory; nevertheless, for non-abelian (compact) $G_0$, it becomes
quantized if the quantum theory is to be well defined ($1/\beta^2
\in {\Bbb Z}^+$)~[\Ref{WITT}].

\section{Theories with a real positive action.}

The above construction of the non-abelian Toda equations makes
perfect sense for any choice of $\{g, \sigma, \Lambda_\pm\}$, without
specifying a particular real form of the Lie group $G_0$. However,
unless $G_0$ is abelian, the compact
real form has to be chosen in order to ensure 
the theory has a positive-definite kinetic term~[\Ref{NOS}]. 
This imposes a reality condition on
the field $h^\dagger = h^{-1}$ that is consistent with the 
equations-of-motion only if $\Lambda_{ \pm}^{ \dagger} = \Lambda_{ \pm}$, which
ensures, moreover, that the potential $V(h)$ is real. Since $\Lambda_+,
\Lambda_{-}^{\dagger} \in g_{\overline{k}}\>$ and $\Lambda_-,
\Lambda_{+}^{\dagger} \in g_{\overline{N-k}}\>$,  the reality condition implies
that $\overline{ 2 k} \>=0$. 

The following cases can be distinguished:

(i) $k=0$ and $N\geq1$. Since $h\in G_0$ and 
$\Lambda_\pm\in g_{\overline 0}$, these theories can be described
just in terms of $g_{\overline 0}$, which in general has the form
$u(1)\oplus\cdots\oplus u(1)\oplus g_{\rm ss}$, where $g_{\rm ss}$ is
semi-simple. However, we shall describe later how the
$u(1)$-fields, associated to the centre of
$g_{\overline 0}$, correspond to flat directions of the potential
and have to be
eliminated if we wish to consider theories with a mass-gap. Therefore,
in this respect, only
the semi-simple part of $g_{\overline 0}$ is important, and, hence, all
the non-equivalent theories of this class can be associated to  
compact semi-simple Lie algebras. The complex sine-Gordon model lies
in this class for $g=su(2)$~[\Ref{CSG},\Ref{PARK2}].

(ii) $N=2$ and $k=1$. In this case, $\sigma$ is an involutive
automorphism of $g$ that induces a gradation of the form
$$
g\> =\> g_{\overline{0}} \> \oplus \> g_{\overline{1}}\>,
\nfr{Two}
where $g_{\overline{0}}$ is a compact subalgebra of $g$. 
This Lie algebra decomposition satisfies the commutation relations
$$
[g_{\overline{0}}\>, \> g_{\overline{0}}] \subset g_{\overline{0}}\>,
\qquad [g_{\overline{0}}\>, \> g_{\overline{1}}] \subset
g_{\overline{1}}\>, \quad {\rm and} \quad [g_{\overline{1}}\>, \>
g_{\overline{1}}]
\subset g_{\overline{0}}\>,
\efr
which implies that these theories are associated to {\it symmetric
spaces} [\Ref{HEL}]. 

These theories are known as the
{\it symmetric space sine-Gordon models\/}~[\Ref{SYMSP},\Ref{PBAK}]
where the symmetric space is of the form $G/G_0$, 
where $G$ is some suitable real form of the group associated to
$g$. However, since $g_{\overline0}$ is compact and $(i\Lambda_{ \pm})^{
\dagger} = - i\Lambda_{ \pm}$ are semi-simple elements of $g_{\overline1}$, 
we will assume that the semi-simple Lie algebra $g$ is also compact, implying
that the relevant symmetric spaces will be of the compact type only.
Nevertheless, we will keep on using the same notation for the compact Lie
algebra $g$ and its complexification. 

It is worth mentioning that all the theories constructed 
in~[\Ref{NOS}] in terms of the integral embeddings of $sl(2)$ into
$g$ are included in this class. They are recovered  by taking
the inner automorphism $\sigma = \exp (\pi i\> {\rm ad}_{ J_0})$
defined by the same Cartan element $J_0$ that specifies the embedding. The
connection of this particular class of theories with symmetric spaces
was originally realized in Ref.~[\Ref{PBAK}].   

(iii) $k>1$ and $N= 2k$. This case appears to be a generalization of the
previous one where now the subalgebra
$\hat{g} = g_{\overline 0}\> \oplus\> g_{\overline k}$ and the
automorphism
$\hat{\sigma} = \sigma^k$ play the role of $g$ and $\sigma$,
respectively. In general, $\hat{g}$ has the form $\hat{g} =
u(1)\oplus \cdots \oplus u(1)\oplus \hat{g}_{\rm ss}$, where
$\hat{g}_{\rm ss}$ is semi-simple and ${\rm cent\/}(\hat{g}) = u(1)
\oplus \cdots \oplus u(1)$ is the centre of $\hat{g}$. However, since the
zero graded generators of ${\rm cent\/}(\hat{g})$ commute with
$\Lambda_\pm$, later we shall see that the associated fields correspond
to flat directions of the  potential and so will be eliminated. In
addition, if $\Lambda_\pm$ have non-vanishing components in ${\rm
cent\/}(\hat{g})$, they only induce a constant term in the
potential~\Pot\ and it is obvious that they do not contribute to the
equations of motion~\NaT. All this shows that only the semi-simple part
of $\hat{g}$ is important and, consequently, that the theories with
$N=2k$ and $k>1$ are already included in the class (ii) by considering
$\hat{g}_{\rm ss}$ and the automorphism $\hat{\sigma} = \sigma^k$, 
which certainly satisfies $\hat{\sigma}^2=I$.

In summary, for non-abelian $g_{\overline
0}$,\note{If $g_{\overline0}$ is abelian, then the restriction to
the compact group does not apply because $G_0$ can then be chosen
to be maximally non-compact, which corresponds to the reality
condition $h^\dagger = h$. Nevertheless, in this case, one does not
expect the theory to have soliton solutions since it is of the
sinh-Gordon rather than sine-Gordon type~[\Ref{HOLL}], and
consequently we will not consider this possibility here.} the only inequivalent
field theories with a positive-definite kinetic term and real potential are
associated to automorphisms such that either $\sigma=I$ (the identity)
or $\sigma^2=I$ (an involution). The first class of theories, 
with $\sigma=I$, involve a compact semi-simple Lie algebra $g= g_{\overline0}$,
and they will be called {\it homogeneous sine-Gordon models} (HSG). The second
class involves the decomposition $g= g_{\overline0} \oplus g_{\overline1}$ of a
compact semi-simple Lie algebra, and corresponds to the {\it symmetric space
sine-Gordon models\/} (SSSG)~[\Ref{SYMSP}] associated to symmetric spaces
$G/G_0$ of compact type.  

So far, the choice of $g$ has been kept as general as
possible. In some cases the
resulting theories can be decoupled into simpler
ones. In particular, an HSG model associated to
a semi-simple Lie algebra $g=g_1\oplus g_2 \oplus\cdots$ can be
decoupled into the set of HSG models associated to each
simple factor $g_1, g_2, \ldots$. Therefore, it will be sufficient to
study the HSG models associated to compact simple Lie
algebras, and a thorough  analysis of these theories will be
presented in a subsequent publication~[\Ref{COMING}].
On the other hand, it follows from the Cartan classification of
compact symmetric spaces that the symmetric space sine-Gordon models
corresponding to a compact Lie algebra $g$ can be decoupled into
SSSG models associated to type~I or type~II symmetric
spaces~[\Ref{HEL}]. Recall that type~I symmetric spaces are associated to a
compact simple Lie algebra $g$ and a involutive automorphism
$\sigma$, while type~II symmetric spaces involve a compact semi-simple Lie
algebra $g=g_1\oplus g_2$, where $g_1 = g_2$ are simple ideals, and a 
involutive automorphism $\sigma$ of $g$ that interchanges $g_1$ and $g_2$.
Notice that it is always consistent to associate a different coupling
constant $\beta$ and mass scale $m$ to each of the decoupled theories. 

\section{Theories with a mass-gap.}

The previous discussion has identified the class
of theories with a positive-definite kinetic term and real potential.
Let us now address the problem of flat directions of the potential. The
potential~\Pot\ exhibits the global symmetry $V(\alpha_- \>h \> 
\alpha_+) =V(h)$ for any $\alpha_\pm$ in the groups $G_\pm$ associated
to the subalgebras ${\rm Ker\/}( {\rm ad\/}_{\Lambda_\pm})\cap
g_{\overline0};$\note{${\rm Ker\/}( {\rm ad\/}_{\Lambda_\pm})$ is the
centralizer of $\Lambda_\pm$ in $g$, {\it i.e.}, the set of $x\in g$ 
that commute with $\Lambda_\pm$.} in other words,
$V(h)$ has
$G_-\times G_+$ (left-right) symmetry. This implies that the quantum theory
corresponding to the action~\Act\ will not have a mass-gap. However,
according to the program of~[\Ref{NOS}], we wish to study theories 
with a mass-gap because such theories are expected to admit an
S-matrix description.  

Recall that $\Lambda_\pm$ are ad-diagonalizable (semi-simple) elements,
which means that the Lie algebra $g$ has the two following
orthogonal decompositions with respect to their adjoint
actions
$$
g\> =\> {\rm Ker\/}( {\rm ad\/}_{\Lambda_\pm}) \> \oplus \>  {\rm
Im\/}( {\rm ad\/}_{\Lambda_\pm})\>.
\efr
Therefore, the equations of motion~\NaT\ imply that those field
configurations such that 
$$
h^{\dagger}\>\partial_+h \in {\rm Ker\/}( {\rm ad\/}_{\Lambda_+})\cap
g_{\overline0}
\qquad {\rm or} \qquad
\partial_-h\> h^{\dagger} \in {\rm Ker\/}( {\rm ad\/}_{\Lambda_-})\cap
g_{\overline0},
\efr
correspond to flat directions of the potential. In order to
remove them, we have to somehow introduce the constraints 
$P_+(h^{\dagger}\>\partial_+h)= P_-(\partial_-h\> h^{\dagger})=0$,
where $P_\pm$ are the projection operators onto the subalgebras ${\rm
Ker\/}( {\rm ad\/}_{\Lambda_\pm}) \cap g_{\overline0}$. 

The way to introduce these constraints was discussed in 
Ref.~[\Ref{PARK1}]. The idea is to gauge a subset of the symmetry
transformations. It is well known that it is not possible to
gauge an arbitrary subgroup of transformations since 
several conditions must be satisfied~[\Ref{GAUGE},\Ref{KIR}]. 
First of all, it
has to correspond to the embedding of some common subgroup
$H$ of $G_\pm$ into $G_-\times G_+$ of the form
$\alpha\mapsto(\alpha_L,\alpha_R)$. In other words, the local
invariance has to be of the form $h\mapsto \alpha_L(x,t)\> h\>
\alpha^{\dagger}_{R}(x,t)$. 
In our case, this first condition implies that, at most, we
will be able to gauge only the transformations generated by the
compact group associated to the subalgebra
$$
\left({\rm Ker\/}( {\rm ad\/}_{\Lambda_+})  \cap {\rm Ker\/}( {\rm
ad\/}_{\Lambda_-}) \right) \cap g_{\overline0}\>,
\efr
which, in general, is not sufficient to remove all the flat directions
of the potential.
Therefore, in order to ensure that this procedure leads to
the required constraints, the two elements $\Lambda_\pm$ have to be
chosen such that
$$
{\rm
Ker\/}( {\rm ad\/}_{\Lambda_+})\cap g_{\overline0}  \>=\> {\rm
Ker\/}( {\rm ad\/}_{\Lambda_-}) \cap g_{\overline0}\>.
\nfr{NewCond}
This additional requirement was not made explicit in~[\Ref{NOS}]
where it is enforced by the stronger condition $\Lambda_+ =
\Lambda_-$. Notice that~\NewCond\ does not imply that $[\Lambda_+,
\Lambda_-]=0$ unless the theory is of the HSG type.

Assuming that $\Lambda_\pm$ satisfy~\NewCond, the flat directions
would be associated to the subalgebra $g_{0}^0 = {\rm
Ker\/}( {\rm ad\/}_{\Lambda_\pm}) \cap g_{\overline 0} $. Then,
$G_+=G_-$ is just the compact group $G_{0}^0$ associated to $g_{0}^0$,
and one has to gauge some set of diagonal transformations of the form
$$
h\mapsto \alpha_L\> h\> \alpha_{R}^\dagger\>, \quad {\rm with}\quad
\alpha_L, \alpha_R \in G_{0}^0\>.
\nfr{Able}
These transformations correspond to an embedding of $G_{0}^0$ into
$G_{0}^0\times G_{0}^0$, which is determined by an embedding of Lie
algebras that can be written as
$$
\eqalign{
g_{0}^0 &\longrightarrow g_{0}^0 \times g_{0}^0\cr
u &\longmapsto (u_L, u_R)\>.}
\efr
The transformations~\Able\ can be gauged if $G_{0}^0$ is an {\it
anomaly free\/} subgroup of $G_{0}^0\times G_{0}^0$, which simply means
that~[\Ref{GAUGE},\Ref{KIR}]
$$
\langle u_L\>, \> v_L\rangle \>=\> \langle u_R\>, \> v_R\rangle,
\nfr{Diag}
for all $u,v$ in $g_{0}^0$. 

The most familiar solution of eq.~\Diag\ is $u_R =u_L$,
which corresponds to the {\it vector} gauge transformations
$h\mapsto \alpha \>h \> \alpha^\dagger$. For our purposes, we will
need to consider more general gaugings. These can be described
by considering a generic automorphism $\tau$ of $g_{0}^0$ and its lift
$\hat{\tau}$ into the group $G_{0}^0$ such that $\hat{\tau}( \exp i \>\phi)
= \exp i\>\tau(\phi)$,  for all $i\phi\in g_{0}^0$. If the choice of
$\tau$ is limited to the set of automorphisms that leave the restriction
of the bilinear form $\langle\>, \>\rangle$ of $g$ to
$g_{0}^{0}$ invariant, $u_R =\tau(u_L)$ is the
general solution of~\Diag. It corresponds to the group of gauge
transformations\note{If
$\widetilde\alpha = \hat\tau(\alpha)$, notice that
${\widetilde\alpha}^{\dagger} \partial_\mu
\widetilde\alpha = \tau\bigl({\alpha}^{\dagger} \partial_\mu
\alpha\bigr)$.} 
$$
h\mapsto \alpha\> h\> \hat{\tau}(\alpha^{\dagger}) \>.
\nfr{GTrans}
These transformations can be gauged by introducing a
gauge field $A_\pm$ taking values in
$g_{0}^0$, and substituting $S_{\rm WZW}[h]$ with the gauged WZW
action $S_{\rm WZW}[h,A_\pm]$ associated to the coset
$G_0/G_{0}^0$. Then, the  gauge invariant action is
$$
S[h, A_\pm]\>=\> {1\over \beta^2} \biggl\{ S_{\rm WZW}[h,A_\pm] 
\>-\>\int d^2 x \>V(h) \biggl\}\>,
\nfr{ActG}
where the gauged WZW is explicitly given by 
$$
\eqalign{
S_{\rm WZW}[h & ,A_\pm]\> = \> S_{\rm WZW}[h]\> + \> {1\over 2\pi}  \int
d^2x \> \Bigl( -\langle A_+\>,\> \partial_-h\>
h^{\dagger}\rangle\cr  
& + \> \langle \tau(A_-)\>, \>
h^{\dagger}\> \partial_+h\rangle\> +\> \langle h^{\dagger}\> A_+\> h\>,\>
\tau(A_-) \rangle\> -\> \langle A_+\>,\>
A_-\rangle\Bigr)\>,\cr}
\nfr{WZW}
and the gauge field transforms as $A_\pm \mapsto \alpha\> A_\pm \>
\alpha^\dagger - \partial_\pm\alpha\>\alpha^\dagger$. The
equations-of-motion, which follow from the variation of~\ActG\ with
respect to $h$, can be written in the zero-curvature form
$$
\Bigl[ \partial_+ \> +\> h^\dagger\>\partial_+h \> +\> h^\dagger \>
A_+\> h\> +\> im\Lambda_+ \>, \> \partial_- \> +\> \tau(A_-)\>+
\> im h^\dagger
\>\Lambda_- h  \Bigr]\> =0\>.
\nfr{GNaT}
On the other hand, the variations with respect to $A_\pm$ lead to the 
constraints
$$
\eqalign{
& P\Bigl( h^\dagger\>\partial_+h \> +\> h^\dagger \>A_+\>
h\Bigr)\> -\> \tau(A_+)\> =\> 0\>, \cr
& P\Bigl( -\partial_- h\> h^\dagger\> +\> h \>\tau(A_-)
h^\dagger \Bigr)\> -\> A_-\> =\> 0\>, \cr}
\nfr{GConst}
where $P$ is the projector onto the subalgebra $g_{0}^0$.
By projecting~\GNaT\ onto $g_{0}^0$ and using~\GConst, one can see that
the gauge field is flat: $[\partial_+ \>+ \>A_+\>, \>\partial_- 
\>+ \>A_-]=0$.

To show that the constraints~\GConst\ actually remove the flat
directions of the potential, it suffices to choose the gauge $A_\pm =0$, 
which is consistent due to the flatness of the gauge field
considered on two-dimensional Minkowski space. In this gauge, the 
equations of motion reduce to the non-abelian Toda equation~\NaT, along
with the constraints
$$
P\left( h^\dagger\>\partial_+h\right)\> =\> P\left( \partial_- h\>
h^\dagger\right)\> =\>0\>.
\nfr{Const}
 
At this stage, it is worth mentioning that, if we restrict the group of gauge
transformations to be of the vector type, the only theories we end up with
correspond exactly with either the deformed coset models of~[\Ref{PARK1}], if
the theory is of the HSG type, or the symmetric space sine-Gordon models worked
out in~[\Ref{PBAK}], if it is a SSSG theory. In contrast, the theories
described by the action~\ActG\ involve a more general group of gauge
transformations. They are associated with $\{g,\sigma,\Lambda_\pm,\tau\}$,
where $g$, $\sigma$, and $\Lambda_\pm$ have to satisfy the constraints
discussed  previously. In the following, we will single out the subclass of
these theories that exhibit a mass-gap. There, the choice of the group of
gauge transformations, {\it i.e.\/}, of
$\tau$, will depend on the field configuration corresponding to the vacuum of
the theory.   

According to~\NaT, the potential~\Pot\ has extrema when 
$$
[\Lambda_+\> , \> h_0^\dagger\> \Lambda_-\> h_0]\>=\>0 \>.
\nfr{Vacuum}
Moreover, since $g_{\overline0}$ is compact, the potential reaches an
absolute minimum for some $x_\pm$-independent field configuration
$h_0$ corresponding to the vacuum, and particles will be
associated to the small fluctations around $h_0$. The success of
the previous procedure for constructing a theory with a mass-gap requires
that all the flat directions of the potential around the vacuum $h_0$
correspond simply to gauge transformations. This means that for any $i\phi,
i\psi \in g_{0}^0$, with $\phi^\dagger =\phi$ and $\psi^\dagger =\psi$,
one can find $i\eta \in g_{0}^0$ such that
$$
{\rm e\/}^{i\phi}\> h_0\> {\rm e\/}^{i\psi} \> =\> {\rm e\/}^{i\eta}\>
h_0\> {\rm e\/}^{-i\tau(\eta)} \>,
\efr
whose linearized form is
$$
h_{0}^\dagger\> \eta\> h_0\> -\> \tau(\eta)\> =\> h_{0}^\dagger\> \phi\>
h_0\> +\> \psi\>.
\nfr{Gold}
Since it has to be possible to solve this last equation for all the
components of $\eta$, the automorphism $\tau$ has to be chosen such that
$$
\tau(u)\> \not =\> h_{0}^\dagger\> u\> h_{0} 
\nfr{VOne}
for any $u$ in $g_{0}^0$. The same conclusion is reached by demanding
that gauge transformations do not leave the vacuum configuration
invariant and, hence, that they do not correspond to flat directions
of the potential around $h_0$. Notice that condition~\VOne\ implies that
it will not be possible to construct massive theories with vector gauge
transformations ($\tau=I$) if the vacuum configuration corresponds just to
$h_0=1$.  

On top of this, there is an additional constraint that the vacuum
configuration $h_0$ has to  satisfy. Let $\bigl(g_{0}^0\bigr)^{\bot}$ be the
orthogonal complement of $g_{0}^0$ in $g_{\overline0}$ with respect to
$\langle\;,\; \rangle$, {\it i.e.\/}
$g_{\overline0}\> =\> g_{0}^0 \oplus \bigl(g_{0}^0\bigr)^{\bot}$,
and let $P$ and $P^\bot$ be the projectors onto $g_{0}^0$
and $\bigl(g_{0}^0\bigr)^{\bot}$, respectively. It is straightforward
to see that eq.~\Gold\ implies 
$$
\eqalignno{
& P\Bigl( h_{0}^\dagger\> \eta\> h_0\Bigr) \> -\> \tau(\eta)\> =\>
P\Bigl( h_{0}^\dagger\> \phi\> h_0\Bigr) \> +\> \psi\>, &
\nameali{PrOne} \cr
& P^\bot \Bigl( h_{0}^\dagger\> \eta\> h_0\Bigr) \> =\>
P^\bot\Bigl( h_{0}^\dagger\> \phi\> h_0\Bigr) \>. &
\nameali{PrTwo} \cr}
$$
If~\VOne\ is satisfied, eq.~\PrOne\ can be solved for all the components
of $\eta$ as functions of $P\Bigl( h_{0}^\dagger\> \phi\> h_0\Bigr) \> +\>
\psi$. Consequently, eq.~\PrTwo\ becomes an identity where the
right-hand-side depends on $\phi$, while the left-hand-side depends on 
$P\Bigl( h_{0}^\dagger\> \phi\> h_0\Bigr) \> +\> \psi$. Therefore, since
$i\phi$ and $i\psi$ are arbitrary elements of $g_{0}^0$, both sides of this
equation have to vanish, which means that
$$
h_{0}^\dagger\> g_{0}^0\>  h_{0}\>= \> g_{0}^0\>. 
\nfr{VTwo}
In other words, $h_0$ has to induce an inner automorphism of
$g_{\overline0}$ that fixes the subalgebra $g_{0}^0$.

The constraint \VOne\ is very restrictive and so the number of
theories with positive and real action that exhibit a mass-gap is
rather limited. Taking into account~\VTwo,
$\rho(u) = h_0\> \tau(u)\> h_{0}^\dagger$ defines an automorphism of $g_{0}^0$,
and eq.~\VOne\ requires that $\rho$ does not leave fixed any element of
$g_{0}^0$. The existence of fixed points under automorphisms of Lie algebras
has been investigated, among other authors, by Borel and Mostow~[\Ref{BM}] and
by Jacobson~[\Ref{JAC}]. In particular, Jacobson proved that any automorphism
of a non-solvable Lie algebra always have a fixed point (Theorem~9
of~[\Ref{JAC}], see also the Theorem~4.5 of~[\Ref{BM}]).\note{Even though the
proof of this important result is quite involved, it is possible to gain some
intuition by considering the subset of inner automorphisms and
of automorphisms that fix a Cartan subalgebra.} However, $g_{0}^0$ is a
reductive Lie algebra, which means that it has the form $u(1)\oplus \cdots
\oplus u(1) \oplus (g_{0}^0)_{\rm ss}$ where
$(g_{0}^0)_{\rm ss}$ is semi-simple. Therefore, the condition that
$\rho$ does not have any fixed point constrains $g_{0}^0$ to be an abelian
subalgebra of $g_{\overline0}$.

The condition that $g_{0}^0$ is abelian implies that the only HSG
theories that have a mass-gap correspond to regular commuting elements
$\Lambda_\pm$ in $g=g_{\overline0}$. This means that $g_{0}^0$ is a Cartan
subalgebra of $g$ and, hence, the massive HSG theories will be associated
with the cosets $G/U(1)^{\times r}$, where $G$ is a semi-simple compact Lie
group of rank $r$.

In contrast, for the SSSG theories, the condition that $g_{0}^0$ is
abelian does not require that $\Lambda_\pm$ are regular elements. In this
case, it is convenient to introduce the notion of a Cartan subspace.
Consider the decomposition $g=g_{\overline0}\oplus g_{\overline1}$
associated with the (compact) symmetric space $G/G_0$. A Cartan subspace
$\ss \subseteq g_{\overline1}$ is a maximal subspace of ad-diagonalizable
(semi-simple) elements which is also an abelian subalgebra of
$g$~[\Ref{HEL},\Ref{VIN}]. For a given symmetric space, all such subspaces have
the same dimension, which defines the rank of the symmetric space. Therefore,
since $g_{0}^0\subset g_{\overline0}$ is an abelian set of semi-simple elements,
it is easy to show that its dimension is bounded as
$$
0\> \leq\> {\rm rank\/}\bigl(G \bigr)\> -\> {\rm rank\/}\bigl(G/ G_0 \bigr)\>
\leq\> {\rm dim\/}\bigl( g_{0}^0 \bigr)\> \leq\> {\rm rank\/}\bigl(G \bigr)\>
-\> 1\>.
\nfr{Bound}
Moreover, in the particular case when $\Lambda_\pm$ are regular
elements and, hence, ${\rm Ker\/}( {\rm ad\/}_{\Lambda_\pm})$ is a
Cartan subalgebra of $g$, the dimension of $g_{0}^0$ equals  the lower
bound ${\rm rank\/}\bigl(G \bigr) - {\rm rank\/}\bigl(G/ G_0 \bigr)$.
In any case, the massive SSSG theories will be associated with cosets
of the form $G_0/U(1)^{\times p}$, where $G/G_0$ is a compact symmetric
space and $p= {\rm dim\/}\bigl( g_{0}^0 \bigr)$. Notice that $p$ may vanish
if the rank of the symmetric space $G/G_0$ equals the rank of $G$ and,
then, the resulting massive SSSG theory is a perturbation of  the WZW model
corresponding to $G_0$.  

Assuming that the previous conditions are satisfied, all the non-gauge
equivalent field configurations around the vacuum are expected to correspond to
massive excitations in the quantum theory. Let us take the $A_\pm =0$ gauge and
$h= h_0 \exp(i\phi) $, with $i\phi \in g_{\overline0}$ and $\phi^\dagger =
\phi$. Then, the linearized constraints~\Const\ and equations-of-motion~\NaT\
are
$$
P(\phi) =0\>, \quad {\rm and} \quad 
\partial_\mu\> \partial^\mu\> \phi \> = \>-\>  4\>
m^2 \> [\Lambda_+\>,\> [h_{0}^\dagger\>
\Lambda_- \> h_0\> , \> \phi]] \>, 
\efr
which show that the {fundamental} particles are associated with the
non-vanishing  eigenvalues of $[\Lambda_+, [h_{0}^\dagger\> \Lambda_- \>
h_0\>,\>  \bullet ]]$ on $g_{\overline0}$. 

Le us introduce a Cartan-Weyl basis for the
(complex) Lie algebra $g$, consisting of a Cartan subalgebra $\hh$ and step
generators $E_{\alb}$, where $\alb$ is a root of $g$. According to the
previous discussion, the Cartan subalgebra can be chosen such that it contains
both $\Lambda_+$ and $\widetilde{\Lambda}_- = h_{0}^\dagger\> \Lambda_- \>
h_0$ (see eq.~\Vacuum), and such that its projection onto $g_{\overline0}$ is
just $g_{0}^0$, $\hh\cap g_{\overline0}= g_{0}^0$. Moreover, it is always
possible to choose the basis of simple roots of $g$, $\{\alb_1, \ldots,
\alb_r\}$, such that $\alb_i\cdot\Lambda_+ \geq0\>\>$,\note{The Cartan
subalgebra is identified with the  root space, which is viewed as a Euclidean
vector space with dimension $r={\rm rank\/}(g)$, and we adopt the vector
notation for the inner product induced there by the bilinear form of $g$.} {\it
i.e.\/}, such that $\Lambda_+$ lies in the fundamental Weyl chamber of $\hh$.
With this choice, $\alb\cdot \Lambda_+$ and $\alb\cdot
\widetilde{\Lambda}_-$ cannot vanish for any root $\alb$ unless the theory is
of the SSSG type and $c\> E_\alb - c^\ast\> E_{-\alb}$ is in $g_{\overline1}$
for any complex number $c$. 

Correspondingly, if $P_0$ is the projector onto $g_{\overline0}$, the massive
excitations are  associated to the generators $P_0(c\> E_\beb -c^\ast\>
E_{-\beb})$ with
$\beb\cdot\Lambda_+\not= 0$ or, equivalently, $\beb
\cdot\widetilde{\Lambda}_- \not= 0$. Their masses are given by
$$
m_{\beb}^2\> =\> 4\>m^2\> (\beb\cdot\Lambda_+) \> (\beb\cdot
\widetilde{\Lambda}_-)\>,
\nfr{Mass}
and they have to be positive because the potential has an absolute minimum at
$h=h_0$. Therefore, whenever $P_0(c\> E_\beb -c^\ast\> E_{-\beb}) \not=0$ for
a positive root $\beb$, we conclude that both $\beb\cdot \Lambda_+$ and
$\beb\cdot \widetilde{\Lambda}_-$ have to be strictly positive.

\chapter{A classification}

According to \Mass, the spectrum of particles states of these
theories is characterised by $\Lambda_+$ and $\widetilde{\Lambda}_-
= h_{0}^\dagger\> \Lambda_-\> h_0$. Therefore, the theories constructed
in the previous section are actually associated with the algebraic data
$\{g,\sigma, \Lambda_+, \widetilde{\Lambda}_-, h_0, \tau\}$, where $g$ is a
semi-simple Lie algebra and $\sigma$ is either the identity ($k=0$), for the HSG
theories, or an involution of $g$ ($k=1$), for the SSSG theories. In the
second place, $\Lambda_+$ and $\widetilde{\Lambda}_-$ are two elements in the
projection of a Cartan subalgebra of $g$ onto $g_{\overline k}$. Their choice
is constrained by the condition that the subalgebra $g_{0}^0 = {\rm Ker} ({\rm
ad\/}_{\Lambda_+}) \cap g_{\overline0} = {\rm Ker}({\rm
ad\/}_{\widetilde{\Lambda}_-}) \cap g_{\overline0}$ is abelian. They
specify the coset $G_0/G_{0}^0$ associated with the gauged WZW action \ActG.
Finally, $h_0$ is a constant element of $G_0$ that conjugates the subalgebra
$g_{0}^0$ into itself, and $\tau$ is an automorphism of $g_{0}^0$ that leaves
the bilinear form of $g$ invariant, and such that $\tau(\bullet) \not=
h_{0}^\dagger (\bullet) h_0$. The field configuration $h_0$ is the vacuum of the
theory and $\tau$ fixes the form of the group of gauge transformations.

However, not all the possible choices lead to non-equivalent theories. Let us
consider the class of theories constructed from a fixed choice of $g$
and $\sigma$, {\it i.e.\/}, the different HSG models corresponding to the same
semi-simple Lie group $G=G_0$, if $\{g, \sigma\}= \{g_{\overline0}, I\}$, or the
SSSG theories associated with a given compact symmetric space $G/G_0$, if $\{g,
\sigma\}= \{g_{\overline0}\oplus g_{\overline1}, \sigma\}$ and $\sigma^2 =I$.
Then, for any constant element $\phi$ in $G_0$, the theories specified by the
data
$$
\{\Lambda_+\>,\> \widetilde{\Lambda}_-\> ,\> h_0\> ,\> \tau\} \quad {\rm and}
\quad \{\phi^\dagger\> \Lambda_+
\>\phi\> ,\> \phi^\dagger\>\widetilde{\Lambda}_-\>\phi\> ,\>
\phi^\dagger\> h_0 \>\phi\> ,\> \tau^\ast\}
\efr
are equivalent if the two groups of gauge transformations are conjugate, {\it
i.e.\/}, if $\tau^\ast (\bullet) = \phi^\dagger\> \tau( \phi\> \bullet \>
\phi^\dagger)
\>
\phi$. This can be easily proved by checking that the former theory is
transformed into the latter through the invertible change of variables
$$
h \mapsto \phi\> h\> \phi^\dagger\>, \quad A_\pm \mapsto \phi\> A_\pm
\> \phi^\dagger\>.
\efr
Therefore, the non-equivalent theories are obtained by restricting the
choice of $\Lambda_+$ and $\widetilde{\Lambda}_-$ to a set of Cartan
subalgebras of $g$ that are non-conjugate under the adjoint action of $G_0$
on $g$. 

It is well known~[\Ref{HEL}] that all the Cartan subalgebras of $g$ are
conjugate by the adjoint action of the Lie group $G$. Therefore, all the
different HSG theories associated to a given semi-simple Lie algebra
$g=g_{\overline0}$ can be recovered by considering all the possible choices of
$\Lambda_+$ and $\widetilde{\Lambda}_-$ in a single Cartan subalgebra of
$g=g_{\overline0}$. 

A parallel result for the non-equivalent SSSG theories corresponding to the same
symmetric space  requires the characterization of the orbits of the Cartan
subalgebras of $g$ under the adjoint action of $G_0$, which is only a
subgroup of $G$. For this reason, in the general case, the number of orbits is
larger than one, and the non-equivalent SSSG theories involve all
the possible choices of $\Lambda_+$ and $\widetilde{\Lambda}_-$ into more than
one non-conjugate Cartan subalgebras. Some results about the $G_0$-orbits of
semi-simple elements of $g_{\overline1}$ can be found in~[\Ref{HEL},\Ref{VIN}]. 

Finally, let us consider the different theories obtained from a given choice of
$\{g,\sigma ,\Lambda_+, \widetilde{\Lambda}_-\}$, which are labelled by
$h_0$ and $\tau$. Although all these theories have the same mass
spectrum, they involve different groups of gauge transformations. Then,
according to the results of [\Ref{KIR}], the  corresponding quantum theories
are expected to exhibit some sort of target-space duality and, hence,
they should be considered as non-equivalent theories. In any case, at the
classical level, the theories associated with $\{h_0, \tau\}$ and $\{h_{0}^\ast,
\tau^\ast\}$, where 
$$
\tau^\ast(\bullet)\> =\> \tau\bigl(h_0\> (h_{0}^\ast)^\dagger\> \bullet \> 
h_{0}^\ast\> h_{0}^\dagger\bigr)\>, 
\nfr{GaugeD} 
can be interchanged through the duality transformation
$$
h \mapsto h_0\> (h_{0}^\ast)^\dagger\> h\>, \quad
A_\pm \mapsto h_0\> (h_{0}^\ast)^\dagger\> A_\pm\> h_{0}^\ast\>
h_{0}^\dagger\>.
\nfr{Dual}
Since the mass spectrum and, hence, $\widetilde{\Lambda}_-$ are kept fixed,
notice that the change
$h_0\mapsto h_{0}^\ast$ also implies that 
$$
\Lambda_-\> =\> h_0\> \widetilde{\Lambda}_-\>
h_{0}^\dagger\> \mapsto\> h_{0}^\ast\> h_{0}^\dagger \> \Lambda_- \>h_0\>
(h_{0}^\ast)^\dagger\>.
\efr
Therefore, if
$h_{0}^\ast\> h_{0}^\dagger \> \Lambda_- \>h_0\> (h_{0}^\ast)^\dagger =
-\Lambda_-$ for some particular choice of $h_0$ and $h_{0}^\ast$, this
change of $\Lambda_-$ is equivalent to $m^2 \mapsto -m^2$. This shows that the
duality transformation \Dual\ is a generalization of the results
of~[\Ref{PARK2}], where a similar transformation in the complex sine-Gordon
model was identified with the Krammers-Wannier duality in the context of
perturbed conformal field theory. Nevertheless, let us remark that the
transformation~\Dual\ is formulated without specifying any particular gauge
fixing prescription, in contrast with the results of~[\Ref{PARK2}].

For a given theory associated with $\{\Lambda_+, \widetilde{\Lambda}_-,h_0,
\tau\}$, eqs.~\GaugeD\ and~\Dual\ also provide a condition for the possibility
of formulating another theory with the same mass spectrum but with vector gauge
transformations. This requires that there exists some alternative vacuum
configuration $h_{0}^\ast$ such that the change $h_0 \mapsto h_{0}^\ast$ implies 
$\tau
\mapsto \tau^\ast =I$. According to \GaugeD, this is only possible if $\tau$ is
the restriction onto $g_{0}^0$ of some inner automorphism of $G_0$.

\chapter{The fields associated to the centre of $g_{\overline0}$}

The fields corresponding to the generators of
${\rm cent\/}(g_{\overline0})$ deserve special attention. First of all,
such fields are only present in the SSSG models.
For simplicity, let us consider the SSSG model corresponding
to a (complex) simple Lie algebra $g$ or, equivalently, to a
symmetric space of type~I, even though it is straightforward to
extend the analysis to the general case of semi-simple algebras (type~II).

Let $iu$ be an element in the centre of $g_{\overline0}$, and decompose the
subalgebra $g_{\overline0}$ as
$$
g_{\overline0}\> =\> {\Bbb R}\>iu\> \oplus\>
\widetilde{g}_{\overline0}\>,
\nfr{CentreD}
where the elements of $\widetilde{g}_{\overline0}$ are orthogonal to
$u$ with respect to $\langle \;,\; \rangle$. Since
$g_{\overline0}$ is compact, the adjoint action of $u$ can be diagonalized
on the complex Lie algebra
$g$, which defines a $u$-dependent integer gradation by adjoint action:
$$
g\> =\> \bigoplus_{j=-M}^M g^{(j)}_u\>, \qquad [u\>, \>a] \>=\>
j\>a\quad {\rm for}\quad a\in g^{(j)}_u\>.
\nfr{GraCe}
Moreover, $u$ is in the centre of $g_{\overline0}$ and, hence, one
has the following inclusions
$$
g_{\overline0}\subset g^{(0)}_u \quad {\rm and} \quad g^{(j)}_u\subset
g_{\overline1}\quad {\rm for \; all}\quad j\not=0\>.
\nfr{Include}
It is always possible to choose a Cartan-Weyl basis for the complex Lie algebra
$g$ such that $u$ is in the fundamental Weyl chamber of the Cartan subalgebra.
Then, if the system of simple roots is $\{\alb_1,\ldots, \alb_r\}$, the
gradation~\GraCe\ is specified by the non-negative integer numbers $s_i =
\alb_i\cdot u$, which give the grade of $E_{\alb_i}$ in the gradation~\GraCe.
The highest root of $g$ is $\thb =\sum_{i=1}^r k_i\> \alb_i$, where $\{k_1,
\ldots, k_r\}$ are the labels of the Dynkin diagram of $g$; hence, in~\GraCe,
$E_{\pm\thb}$ is an element of the subspace $g^{(\pm M)}_u$ with $M=\sum_{i=1}^r
k_i\> s_i$. The step operators $\{E_{\pm \alb_1}, \ldots, E_{\pm
\alb_r}\}$ generate the Lie algebra $g$, and, taking into account
eq.~\Include\ and
$[g_{\overline1}\>,\> g_{\overline1}] \subset g_{\overline0}$, one
is led to the conclusion that $s_i$ can be non-vanishing for a single simple
root $\alb_j$ such that $k_j=1$. In this gradation, $s_i=\delta_{i,j}$ and
$$
g\> = \> g^{(-1)}_u\>\oplus\> g^{(0)}_u\> \oplus\> g^{(1)}_u\>, 
\quad{\rm with}\quad 
g_{\overline0}\subset g^{(0)}_u \quad {\rm and}
\quad  g^{(\pm1)}_u\subset g_{\overline1}\>.
\nfr{Gra}
Using the last equation and taking into account $\Lambda_{
\pm}^\dagger = \Lambda_{\pm}$, the elements $\Lambda_\pm$ can be
decomposed as
$$
\Lambda_{\pm}\>= \> \Lambda_{\pm}^{(-1)}\> +\> \Lambda_{\pm}^{(0)}\>
+ \> \Lambda_{\pm}^{(1)}\>,
\nfr{LamD}
where $\bigl(\Lambda_{\pm}^{(0)}\bigr)^\dagger =
\Lambda_{\pm}^{(0)}$ and $\bigl(\Lambda_{\pm}^{(\pm1)}\bigr)^\dagger =
\Lambda_{\pm}^{(\mp1)}$.

According to~\CentreD, let us consider the field configuration
$$
h\> = \> \widetilde{h}\> \exp (i\>\varphi\> u)\>,
\nfr{Field}
where $\widetilde{h}$ is a field taking values in the compact group
associated to $\widetilde{g}_{\overline0}$, and
$\varphi=\varphi(x,t)$ is the field associated to $u$. For
simplicity, we will assume that
$P(u)=0$ and consider the equations of motion in the $A_\pm =0$
gauge. Then, the non-abelian affine Toda equation~\NaT\ yields two
decoupled equations
$$
\eqalign{
&\partial_-\bigl(\widetilde{h}^\dagger  \>
\partial_+\widetilde{h}\bigr)\> =\>-\> m^2\> [\Lambda_+\> ,\> 
\widetilde{h}^\dagger\> \Lambda_{-}\> \widetilde{h}]\cr
&\qquad -\> m^2\> (\cos\varphi\> -\> 1) \Bigl([\Lambda_{+}^{(1)}\> ,\>
\widetilde{h}^\dagger\> \Lambda_{-}^{(-1)}\> \widetilde{h}]\>
+\> [\Lambda_{+}^{(-1)}\> ,\> \widetilde{h}^\dagger\> 
\Lambda_{-}^{(1)}\> \widetilde{h}]\Bigr)\>, \cr}
\nfr{FieldsO}
and
$$
\partial_+\partial_-\> \varphi\> = \> -\>{2\> m^2\over \langle u,
u\rangle} \> \sin\varphi\> \langle \Lambda_{+}^{(-1)}\> , \>
\widetilde{h}^\dagger\> \Lambda_{-}^{(1)}\> \widetilde{h}\rangle\>,
\nfr{FieldC}
which show that the equations-of-motion admit a reduction,
preserving integrability, by taking $\varphi(x,t)=0$. Since this
result applies to a generic element $u$ in the centre of
$g_{\overline0}$, it implies that all the fields associated to the
centre of $g_{\overline0}$ can be decoupled whilst preserving
integrability.

Notice that the analysis leading to~\Gra\ also shows that the dimension of the
centre of $g^{(0)}_u$ is 1, which means that ${\rm cent\/}(g^{(0)}_u)\> =\>
{\Bbb R\/}\> iu$. Nevertheless, eq.~\Gra\ only implies that
${\rm cent\/}(g^{(0)}_u) \subset {\rm cent\/}(g_{\overline0})$, {\it i.e.}, the
centre of $g_{\overline0}$ is not one-dimensional in the general case. 
However, if the identity $g_{\overline0} =g^{(0)}_u = {\rm Ker\/}({\rm
ad}_u) $ is satisfied for some $u$ in ${\rm
cent\/}(g_{\overline0})$, one can ensure that, in this particular case,
the dimension of the centre of $g_{\overline0}$ is actually 1. For instance,
this is the case of the theories constructed in~[\Ref{NOS}] from the
integral embeddings of  $sl(2)$ into $g$. There, if $J_0$ is the Cartan
element of the embedded $sl(2)$ subalgebra, $g_{\overline 0} = {\rm
Ker\/}({\rm ad \/}_{J_0} ) = g^{(0)}_{J_0}$ and $J_0$
spans the one-dimensional centre of $g_{\overline 0}$. Moreover,
in~[\Ref{NOS}], $\Lambda_+ = \Lambda_- = J_+ +J_-$, which implies that
$$
\Lambda_{\pm}^{(0)}\> =\> 0\>, \quad \Lambda_{\pm}^{(1)}\> =\> J_+ \>,
\quad{\rm and}\quad
\Lambda_{\pm}^{(-1)}\> =\> J_-\>.
\efr
This shows that eq.~\FieldsO\ has the solution $\widetilde{h} =1$
and, correspondingly, eq.~\FieldC\ becomes the sine-Gordon equation
for $\varphi$. Therefore, if $\varphi$ is not decoupled, the theories
of~[\Ref{NOS}] describe the interaction between some set of
non-abelian Toda fields corresponding to $\widetilde{h}$, and the
sine-Gordon field $\varphi$.  

\chapter{Parity invariant theories}

As pointed out in Ref.~[\Ref{NOS}], it is of particular interest to
consider the class of theories that exhibit parity invariance. In
particular, we require that the parity transformation fixes the vacuum.

By analysing the equations-of-motion~\GNaT, one can check that the theory has
the symmetry $x\mapsto-x$, or $x_\pm \mapsto x_\mp$, 
along with $h\mapsto h_0\> h^\dagger\>
h_{0}$, $A_+\mapsto h_{0}\> \tau(A_-)\> h_{0}^\dagger$, and $A_-\mapsto 
\tau^{-1}(h_{0}^\dagger\> A_+\> h_0)$, only if
$$
\widetilde{\Lambda}_-\> =\> h_{0}^\dagger\> \Lambda_-\> h_{0}\> = \> \mu\>
\Lambda_+
\>\equiv\>
\mu\> \Lambda\>,
\nfr{Parity}
where $\mu$ is some real number. Moreover, the constraints~\GConst\ have to be
invariant under this transformation, which leads to the following relation
$$
\tau\bigl(h_0\> \tau(A_\pm)\>h_{0}^\dagger \bigr)\> =\> h_{0}^\dagger
\> A_\pm\> h_0 \>. 
\efr
It has to be satisfied independently of the
particular values of the components of $A_\pm$, which are
functionals of $h$. Therefore, taking into account~\VTwo, this means that
the automorphism
$\rho(\bullet) \>= \> h_0\> \tau(\bullet)\> h_{0}^\dagger$ of $g_{0}^0$
has order two, $\rho^2 =I$. Consequently, $\rho$ can be
diagonalized on $g_{0}^0$ with eigenvalues $\pm 1$, but the
condition~\VOne\ implies that $\rho= -I$. 

Consequently, in the parity invariant theories, the
automorphism $\tau$ is given by
$$
\tau(u) \>=\> - \> h_{0}^\dagger \> u \> h_{0} \quad {\rm for
\; all}\quad u \in g_{0}^0\>,
\nfr{VThree}
which indicates that the group of gauge transformations is completely specified
by the vacuum configuration. Moreover, let us point out that the
condition that $\rho=-I$ is an automorphism of $g_{0}^0$ would constrain by
itself the subalgebra $g_{0}^0$ to be abelian, as can be easily checked
by considering the identities   
$$
\eqalign{
\rho \bigl( [u&\>, v\>] \bigr) \> =\> -\> [u\>, v\>]\cr
& =\> [\rho(u) \>,\> \rho(v)] \> =\> [-u\> ,\> -v]\> =\> [u\>, v\>]\>,\cr}
\nfr{Abelian}
for any $u,v \in g_{0}^0$. 

Taking into account~\Mass\ and~\Parity, the
mass of the particle associated with a root $\beb$ is given by
$m_{\beb}^2\> =\>  4\>m^2\>\mu\> \bigl(\beb\cdot\Lambda\bigr)^2$. Since they
have to be positive, $\mu$ is a positive number that can be fixed to
$\mu=+1$.

Therefore, parity invariant theories can be constructed in terms of
a semi-simple Lie algebra $g$ as follows. Let $\sigma$ be an
automorphism of $g$ such that either $\sigma=I$ and $k=0$ (HSG), or
$\sigma^2=I$ and $k=1$ (SSSG), and let us choose a ad-diagonalizable
element $\Lambda=\Lambda^\dagger \in g_{\overline k}$ such that
$g_{0}^0 = {\rm Ker}({\rm ad\/}_\Lambda) \cap g_{\overline 0}$ is
abelian. Then, for any element $h_0 \in G_0$ inducing an inner
automorphism of $g_{\overline0}$ that fixes $g_{0}^0$, the theory is
defined by the action~\ActG\ with the potential
$$
V(h)\> =\> -\>{m^2\over 2\pi} \> \langle  \Lambda\>,\> h^{\dagger}\>
\bigl( h_0\> \Lambda \> h_{0}^\dagger\bigr)\> h \rangle \>,
\nfr{PotP}
whose gauge symmetry is specified by the automorphism
$\tau$ given by~\VThree. In this case,
$\hat{\tau}(
\alpha) = h_{0}^\dagger \alpha^\dagger h_0$, and the resulting group
of gauge transformations is~\note{If $\alpha=\exp i\phi$,
with $i\phi\in g_{0}^0$, then $\hat{\tau}( \alpha) = \exp
\>i\tau(\phi) = \exp -i(h_{0}^\dagger
\phi h_0) = h_{0}^\dagger \alpha^\dagger h_0\>$.}
$$
h\mapsto \alpha\> h\>  \hat{\tau}(\alpha^\dagger)\>
=\> \alpha\> h\> \bigl(h_0^\dagger\> \alpha \> h_{0}\bigr)
\quad {\rm for \; any}\quad \alpha \in G_{0}^0\>;
\nfr{GaugeP}
the corresponding transformation of the vacuum configuration is $h_0
\mapsto \alpha^2 \> h_0$. This theory is invariant under the parity
transformation
$$
x\mapsto -x\>, \quad h\mapsto h_0\> h^\dagger \> h_0\>, \quad {\rm and}\quad
A_\pm \mapsto -\>A_\pm\>.
\nfr{Parity}
Therefore, parity invariant theories can be labelled by the data $\{g,
\sigma, \Lambda, h_0\}$, which, using the terminology of Section~3,
correspond to $\Lambda_+ = \widetilde{\Lambda}_- =\Lambda$ and $\tau(\bullet)
=-h_{0}^\dagger (\bullet) h_0$. 

The massive excitations are associated to those roots $\beb$ of
$g$ such that
$\beb\cdot\Lambda \not= 0$, and their masses are given by
$$
m_{\beb}\> =\>  2\>m\> |\beb\cdot\Lambda|\>.
\nfr{MassP}
As explained in Section~2.2, it is always possible to choose a basis
of simple roots $\{\alb_1,\ldots, \alb_r\}$ of $g$ such that
$\Lambda$ is in the fundamental Weyl chamber of the Cartan subalgebra,
{\it i.e.\/}, $\alb_i \cdot\Lambda \geq0$ for all $i=1, \ldots, r$.
With this choice, the condition that $g_{0}^0$ is abelian means that
$\alb_i \cdot\Lambda =0$ only if $P_0(c\>E_{\alb_i}\> -\> c^\ast \>E_{-
\alb_i})=0$. With this choice, let $I$ be the set of integer numbers such that
$\alb_j \cdot\Lambda \not=0$ for $j\in I$. Then, since any positive root is of
the form $\beb = \sum_{i=1}^r n_i\alb_i$ for some non-negative
integers $n_i$, the mass of the fundamental particle associated to
the roots $\pm \beb$ is
$$
m_{\beb}\> =\>  \sum_{j\in I}\> n_j \> m_{\alb_j}\>.
\efr
This suggests that solitons corresponding to non-simple roots are
bound-states at threshold of solitons associated to simple roots. An
identical phenomenon occurs for monopoles in $N=4$ supersymmetric
gauge theories [\Ref{GENM}].

\chapter{Some general properties}

At this stage, let us make some 
comments about gauge fixing. The
action~\ActG\ describes a theory that is invariant under the 
gauge transformations~\GTrans. Hence, it is pertinent to ask
about the possible gauge fixing prescriptions.
In this regard, there are two
particular useful choices. The first one will be called the {\it local}
gauge fixing prescription and consists in choosing some canonical
form $h^{\rm can}$ such that any $h$ can be taken to that form by
means of a non-singular $h$-dependent gauge transformation.
Therefore, for any $h\in G_0$ there exist two local functionals
$i\phi^{\rm can}[h] \in g_{0}^0$ and
$h^{\rm can}[h]$ such that
$$
h\> =\> \exp \bigl(i\phi^{\rm can}[h]\bigr)\>\> h^{\rm can}[h] \>\>
\exp \bigl(-i\tau(\phi^{\rm can}[h])\bigr)\>.
\nfr{GSlice}
Moreover, under a gauge transformation
$h\mapsto h' = {\rm e\/}^{iu}\> h\> {\rm e\/}^{ -i\tau(u)}$, with
$iu$ taking values in $g_{0}^0$, $h^{\rm can}[h]$ is gauge invariant,
$h^{\rm can}[h'] =h^{\rm can}[h]$, while, since $g_{0}^0$ is abelian, 
$\phi^{\rm can}[h]$ transforms as
$$
\phi^{\rm can}[h'] \>=\> \phi^{\rm can}[h]\> +\> u\>.
\efr
The local
gauge fixing prescription is simply $\phi^{\rm can}[h]=0$, and solving
the constraints~\GConst\ for $A_\pm$ as local functionals of $h$
allows one to obtain a local gauge-invariant action: $S^{\rm
can}[h]\equiv S[h,A_\pm[h]]$.

The second gauge fixing prescription will be called the 
{\it Leznov-Saveliev} (LS) prescription. It consists in choosing
$A_\pm=0$, which can be done due to the on-shell flatness of the gauge
field considered on two-dimensional Minkowski space; notice that this
condition does not fix the global gauge transformations. In the
LS gauge, the equations of motion~\GNaT\ reduce to the
non-abelian Toda equation~\NaT, and the constraints~\GConst\ reduce 
to~\Const. These constraints, as pointed out in Ref.~[\Ref{PARK1}], cannot be 
solved locally. Nevertheless, using the method of Leznov and Saveliev, the
explicit general solution of the non-abelian Toda equation along with
the constraints~\Const\ can be obtained using the
representation theory of affine Kac-Moody algebras~[\Ref{LS}]. Actually, it 
is particularly easy, in this gauge, to obtain the multi-soliton solutions by
means of the so-called {\it solitonic specialization}~[\Ref{OLIVE}].
Moreover, many of the relevant calculations involving solitons, such as
the calculation of their  masses~[\Ref{LUIZ}] and scattering time-delays,
are greatly simplified. 

In these theories, there exist conserved charges. In order to uncover
them, recall that the potential has the $G_{0}^0\times G_{0}^0$ symmetry
property $V({\rm e}^{  i\phi}\> h\>{\rm e}^{  i\psi}) =V(h)$ for any
$i\phi, i\psi\in g_{0}^0$, with $\phi^\dagger=\phi$ and $\psi^\dagger=
\psi$. Taking into account this, and requiring that the transformations fix
the vacuum $h_0$, one can check that the theory exhibits an abelian global
symmetry with respect to the tranformations
$$
h\mapsto \alpha\> h\> (h_{0}^\dagger\> \alpha^\dagger \>
h_0)\> , \quad A_\pm \mapsto A_\pm\>,  
\nfr{AbSym}
for each element $\alpha$ in the compact abelian group $G_{0}^0$. 

Along with this abelian global symmetry there is a continuity equation 
$$
\bigl[ \partial_+\> +\> A_+\>,
\>\partial_-\> +\> A_-\bigr]\>=\> 0\>,
\efr
which is nothing else than the on-shell flatness condition for the gauge
fields. Then, taking into account the gauge transformations of
$A_\pm$ and of $\phi^{\rm can} =\phi^{\rm can}[h]$, defined in~\GSlice, the
corresponding gauge invariant conserved Noether current is
$$
J^\mu \> = \> \epsilon^{\mu\> \nu}\> \bigl( A_\nu\> +\> i \partial_\nu
\>\phi^{\rm can} \bigr)\>.
\nfr{Noether}
It is important to remark that, in the local gauge, $J^\mu$ and the
associated conserved Noether charge $Q  = \> \int_{-\infty}^{+\infty}
dx\> J^0 $ are local functionals of $h$.

Finally, let us briefly study the energy-momentum tensor of the
theory described by the action~\ActG. Its components are
$$
\eqalignno{
T_{+\>+}\> &= \> -\>{1\over 8\pi\beta^2 }\> \langle
\partial_+h h^\dagger\>,\>   \partial_+h h^\dagger\> +\> 2\>
A_+\rangle \cr 
T_{-\>-}\> &= \> -\> {1\over 8\pi\beta^2 }\> \langle
h^\dagger\>\partial_-h \> ,\> h^\dagger\>\partial_-h \> -\> 2\>
\tau(A_-)\rangle \cr
T_{+\>-}\> &= \>- \> { m^2\over 4\pi\beta^2 }\> \langle
\Lambda_+\>, \> h^\dagger\> \Lambda_- \> h\rangle \>, &
\nameali{Tensor}\cr}
$$
and it can be checked that it is explicitly gauge invariant. In the LS
gauge, and using the solitonic specialization of the Leznov and Saveliev
solution~[\Ref{OLIVE}] and the results of~[\Ref{LUIZ}], the calculation of the 
energy and momentum carried by solitons is greatly simplified,  and their
relation to the boundary conditions of the solitons can be clarified. Actually,
in this gauge, eq.~\Noether\ shows that the values of the abelian conserved
Noether charges are also explicitly related to the boundary conditions
$$
Q \> = \> \int_{-\infty}^{+\infty} \> dx\> J^0\> =\>
\>-i\> \phi^{\rm can} \Bigm|_{-\infty}^{+\infty}\>.
\efr

\chapter{Discussion}

We have constructed two series of relativistic two-dimensional field theories
whose equations-of-motion are related to the non-abelian Toda equations, namely
the Symmetric Space (SSSG) and the Homogeneous (HSG) sine-Gordon models. They
are singled out because they can be described by an action with a
positive-definite kinetic term and a real potential, and because they also
exhibit a mass-gap. The action consists of the gauged WZW action of
a coset model plus a potential, which manifests their interpretation as
perturbed conformal field theories. Moreover, the constructed theories are
classically integrable, admit soliton solutions, and exhibit internal
symmetries. Therefore, we expect that the semi-classical quantization of their
soliton solutions will give rise to a spectrum consisting of massive charged
particles.

The SSSG theories are associated to compact symmetric spaces, while the HSG
theories involve compact semi-simple Lie groups. A compact symmetric
space $G/G_0$ is related to a compact semi-simple Lie algebra $g$ and an
involutive automorphism $\sigma$ that induces the decomposition $g=
g_{\overline0}\oplus g_{\overline1}$ into the eigenspaces where the eigenvalue 
of $\sigma$ is~$+1$ and~$-1$; $G$ and $G_0$ are the compact Lie groups whose Lie
algebras are $g$ and $g_0$, respectively. If we also let $G_0$ and
$g_{\overline0}$ denote a compact semi-simple Lie group and its Lie algebra, we
can give a joint description of the data needed to define both series of
theories, where, using this notation, the field always takes values in $G_0$.
For a given compact symmetric space $G/G_0$ (compact semi-simple Lie group
$G_0$), the different SSSG (HSG) theories are labelled by the data
$\{\Lambda_+, \widetilde{\Lambda}_-, h_0, \tau\}$. The first two,
$\Lambda_+$ and $\widetilde{\Lambda}_- $, are two semi-simple elements of
$g_{\overline1}$ ($g_{\overline0}$) such that their centralizer in
$g_{\overline0}$ is an abelian subalgebra $g_{0}^0$. The choice of $\Lambda_+$
and $\widetilde{\Lambda}_- $ determines the form of the potential, the
mass-spectrum, and the coset $G_0/G_{0}^0$ corresponding to the
gauged WZW term in the action, where $G_{0}^0$ is of the form $U(1)^{\times
p}$. The precise form of the group of gauge transformations is specified by
an automorphism $\tau$ of $g_{0}^0$ that preserves the bilinear form of $g$.
Theories with the same spectrum but different groups of gauge transformations
are expected to be related by target-space dualities~[\Ref{KIR}]. 

The choice of the group of gauge transformations is constrained by the
condition that it has to allow one to eliminate all the flat directions of the
potential, which means that $\tau$ is related to the vacuum configuration
$h_0$. Furthermore, the same condition requires that the adjoin action of $h_0$
on $g_{0}^0$ has to be an automorphism, which constrains the possible vacuum
configurations. Therefore, different vacuum configurations imply different
groups of gauge transformations and, because of this, we consider $h_0$ as an
additional data. From this description, it is apparent that the SSSG theories
associated to $G/G_0$ can also be viewed as the reduction of the HSG models
related to $G$. Moreover, we have also described other reductions of the SSSG
theories that maintain integrability. They consists in decoupling the fields
associated to the centre of $G_0$.

The resulting theories are generalizations of
the sine-Gordon~[\Ref{SG}] and complex sine-Gordon
theories~[\Ref{CSG}] and so we expect 
the spectrum of quantum states can be understood in terms of
the semi-classical quantization of the soliton and other lump-like solutions.
However, since in general the field takes values in a non-abelian group, an
important difference between these theories and the sine-Gordon theory is that
the coupling constant $\beta$ is quantized at the quantum level ($1/\beta^2 \in
{\Bbb Z}^+$), something that has already been observed in the complex
sine-Gordon theory~[\Ref{CSG}]. 

The important question of the quantum
integrability of these theories can be addressed by considering their
description as perturbed conformal field theories. The existence of
quantum conserved charges can then be investigated by using the method of
Zamolodchikov~[\Ref{ZAM}]. 
The next stage of analysis involves trying to establish
the form for the exact S-matrix for the scattering of the soliton and
particle states. A powerful constraint on the soliton S-matrix arises
from taking the semi-classical limit leading to a relation with the time
delays that occur in the classical scattering [\Ref{JW}].
The time-delays themselves can
be easily extracted from the solitonic
specialization of Leznov and Saveliev~[\Ref{OLIVE}].

\bjump
\centerline{\bf Acknowledgements}

\noindent
C.R.F.P., M.V.G. and J.L.M. would like to thank Joaqu\'\i n~S\'anchez
Guill\'en and Alfonso~V. Ramallo for useful discussions. They are supported
partially by CICYT (AEN96-1673) and DGICYT (PB93-0344).
T.J.H. is supported by a PPARC Advanced Fellowship.

\references

\beginref
\Rref{ZZ}{A.B. Zamolodochikov and Al. B. Zamolodchikov,
Ann. Phys. {\bf120} (1979) 253}
\Rref{BM}{A.~Borel and G.D.~Mostow, Annals of Math.~{\bf 61} (1955)
389-405.}
\Rref{JAC}{N.~Jacobson, Pacific J. Math.~{\bf 12} (1962) 303-315.}
\Rref{ZAM}{A.B.~Zamolodchikov, Int. J. Mod. Phys. {\bf A3} (1988) 743.}
\Rref{VIN}{E.B.~Vinberg, Math. USSR-Izv. {\bf 10} (1976) 463.}
\Rref{SG}{R.F. Dashen, B.~Hasslacher, and A.~Neveu, Phys. Rev. {\bf D 11}
(1975) 3424.}
\Rref{NAT}{J.~Underwood, {\sl Aspects of Non-Abelian Toda Theories\/},
Imperial/TP/92-93/30, hep-th/9304156.}
\Rref{WITT}{E.~Witten, Commun. Math. Phys. {\bf 92} (1984) 455.}
\Rref{NOS}{T.J.~Hollowood, J.L.~Miramontes, and Q-H.~Park, Nucl.
Phys. {\bf B 445} (1995) 451.}
\Rref{CSG}{N.~Dorey and T.J.~Hollowood, Nucl. Phys. {\bf B 440} (1995)
215.}
\Rref{HEL}{S.~Helgason, {\sl Differential geometry, Lie groups and
symmetric spaces}, (Academic Press, New York, 1978).}
\Rref{SYMSP}{R.D'Auria, T. Regge, and S. Sciuto, Phys. Lett.  {\bf 89
B} (1980) 363; Nucl. Phys. {\bf 171 B} (1980) 167;\newline
H.~Eichenherr and K.~Polmeyer, Phys. Lett. {\bf 89 B}
(1979) 76; H.~Eichenherr, Phys. Lett. {\bf 90 B} (1980) 121;\newline
V.E.~Zakharov and A.V.~Mikhailov, Sov. Phys. JETP {\bf 47} (1978) 1017.}
\Rref{PBAK}{I.~Bakas, Q-H.~Park, and H-J.~Shin, Phys. Lett. {\bf B 372}
(1996) 45.}
\Rref{COMING}{C.R.~Fern\'andez-Pousa, M.V.~Gallas, T.J.~Hollowood, and
J.L.~Miramontes, work in preparation.}
\Rref{BAKAS}{I.~Bakas, Int. J. of Mod. Phys. {\bf A 9} (1994) 3443.}
\Rref{PARK1}{Q-H.~Park, Phys. Lett. {\bf B 328} (1994) 329.}
\Rref{GAUGE}{E.~Witten, Commun. Math. Phys. {\bf 144} (1992) 189.}
\Rref{KIR}{E.B.~Kiritsis, Mod. Phys. Lett. {\bf A 6} (1991) 2871.}
\Rref{PARK2}{Q-H.~Park, and H-J.~Shin, Phys. Lett. {\bf B 359} (1995)
125.}
\Rref{LS}{A.N.~Leznov and M.V.~Saveliev, Commun. Math. Phys. {\bf
89} (1983) 59; {\sl Group theoretical methods for integration of
non-linear dynamical systems\/}, Prog. Phys.~15 (Birkhauser, Basel,
1992).}
\Rref{OLIVE}{D.I.~Olive, N.~Turok and J.W.R.~Underwood, Nucl. Phys.
{\bf B 401} (1993); Nucl. Phys. {\bf B 408} (1993) 565; \newline
D.I.~Olive, M.V.~Saveliev and J.W.R.~Underwood, Phys. Lett. {\bf
B 311} (1993) 117.}
\Rref{LUIZ}{L.A.~Ferreira, J.L.~Miramontes, and J.~S\'anchez
Guill\'en, Nucl. Phys. {\bf B 449} (1995) 631.}
\Rref{HOLL}{T.J.~Hollowood, Nucl. Phys. {\bf B 384} (1992) 523.}
\Rref{GENM}{J.P. Gauntlett and D. A. Lowe, `Dyons and S-duality in 
$N=4$ Supersymmetric Gauge Theory', {\tt hep-th/9601085}\newline
K. Lee, E.J. Weinberg and P. Yi, `Electromagnetic Duality
and SU(3) Monopoles', {\tt hep-th/9601097}, `The Moduli Space
of Many BPS Monopoles for Arbitrary Gauge Groups', {\tt
hep-th/9602167}}
\Rref{JW}{R. Jackiw and G. Woo, Phys. Rev. {\bf D12} (1975) 1643}
\endref   

\ciao